\documentclass[preprint]{llncs}

\usepackage{latexsym}
\usepackage{paralist}
\usepackage{amssymb}
\usepackage{soul}
\usepackage{xcolor}
\usepackage{enumitem}
\usepackage{mathtools}
\usepackage[all]{xypic}

\usepackage{paralist}

\usepackage{graphics}

\newcommand{\Q}{\mathbb{Q}}  
\newcommand{\C}{\mathbb{C}}
\newcommand{\R}{\mathbb{R}}
\newcommand{\F}{\mathbb{F}}
\newcommand{\Z}{\mathbb{Z}}
\newcommand{\N}{\mathbb{N}}
\newcommand{\quat}{\mathbb{H}}

\usepackage{amssymb}

\newtheorem{exam}[example]{Example}
\newtheorem{lem}[example]{Lemma}
\newtheorem{propos}[example]{Proposition}
\newtheorem{cor}[example]{Corollary}
\newtheorem{defin}[example]{Definition}

\newtheorem{principle}[example]{Principle}
\newcommand{\id}{\mathrm{id}}

\begin{document}

\begin{frontmatter}
\title{\LARGE A model of systems with modes and mode transitions}
\author{Edwin Beggs  \and John V.~Tucker}
\institute{College of Science, Swansea University, Wales, U.K. }
\titlerunning{A model of systems with modes}
\authorrunning{Edwin Beggs and John V.~Tucker}
\maketitle
\pagenumbering{arabic}
\begin{abstract}
We propose a method of classifying the operation of a system into finitely many modes. Each mode has its own objectives for the system's behaviour and its own mathematical models and algorithms designed to accomplish its objectives.  A central problem is deciding when to transition from one mode to some other mode, a decision that may be contested and involve partial or inconsistent information or evidence. We model formally the concept of modes for a system and derive a family of data types for analysing mode transitions. The data types are simplicial complexes, both abstract and realised in euclidean space $\mathbb{R}^{n}$.   In the data type, a mode is represented by a simplex. Each state of a system can be evaluated relative to different modes by mapping it into one or more simplices. This calibration measures the extent to which distinct modes are appropriate for the state and can decide on a transition. We explain this methodology based on modes, introduce the mathematical ideas about simplicial objects we need and use them to build a theoretical framework for modes and mode transitions. To illustrate the general model in some detail, we work though a case study of an autonomous racing car.
\end{abstract}
\end{frontmatter}

\section{Introduction} 

Consider a complex system that is designed to have various modes of operation and behaviour. We suppose that each mode is based on a particular set of objectives for the system's operation and will have a particular set of mechanisms  -- mathematical equations and algorithms -- designed to accomplish these objectives.  Thus, modes have a certain independence from one another and the complex system \textit{is} the combination of its set of modes. A central problem for a system with modes is to decide when it is necessary to transition from one mode to some other mode. The system is dependent on its internal state and on data it receives from monitoring its environment. Decisions to change modes may need to be made in the presence of partial or competing information. 

To illustrate, think of an autonomous car whose basic modes on a race track are: start, drive, corner, stop;  these modes may be refined for dry and wet weather conditions, enhancing the drive and corner modes making six modes. Off the race track the number of modes multiply: new basic modes are needed for turn right, turn left, reverse, and park; and, most important, alert modes are needed for when the presence of traffic signals, nearby cars and pedestrians are detected and need interpretation, leading to emergency modes for exceptional situations. Modes can overlap and be organised into a sort of hierachy.

In this paper we will explore theoretically the idea of modes of operation and mode transitions.  Starting from scratch, with simple examples, e will create an algebraic framework for decomposing and recomposing the behaviour of a system using modes. To do this we develop a family of mathematical models of component concepts, as follows:

\begin{enumerate}
\item \textit{Modes}: An algebraic structure to model a finite set of modes and mode transitions.

\item \textit{Data}: A state space and algebraic data type that are specific to each mode, which are needed for computations with data monitoring the environment;

\item \textit{System Architecture}: An algebraic structure to assemble these mathematical components, indexed by the modes, to model the structure of the whole system.

\item \textit{Quantification}: Geometric data types to quantify and visualise the modes, and the semantic complexities involved in mode transition decisions.

\end{enumerate}

\begin{figure}
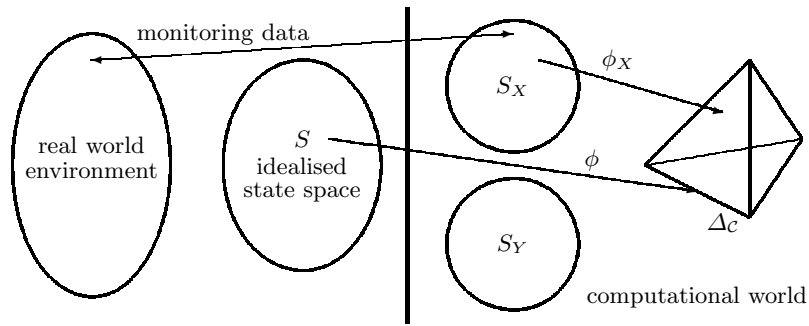

\begin{center}

\unitlength 0.7 mm

\caption{The different worlds of the control system}
\label{basicdiag}
\end{center}
\end{figure}

With the help of simple examples, we develop formal notions about modes and the role of data in changing modes. We propose that the algebraic structure of a family of modes for a system can be represented by an \textit{abstract simplicial complex} $\mathcal{C}$ in which the modes are indexed by the abstract simplices $X \in \mathcal{C}$. The algebra of the simplices provide a self-contained hierarchical structure for specifying, comparing and refining modes. Simplicial maps that map simplices to simplices provide an ability to compare, calibrate, expand or retract models of modes.

To each mode, indexed by a simplex $X \in \mathcal{C}$, we associate a package $D_X$ containing state space $S_X$ and a data type $\mathcal{A}_X$, both of which are local to the mode $X$. The space $S_X$ contains the possible monitoring data needed by the mode $X$, and the data type $\mathcal{A}_X$ contains all the operations and tests needed for the algorithms $\mathcal{B}_X$ that control the system when in that mode. 

In Figure~\ref{basicdiag} we set out the basic architecture. We can begin by imagining an idealised state space $S$ of the whole system in physical terms.  The designers' original ideas of the splitting of the system into modes is idealised by a map $\phi$ from $S$ into the
 simplicial complex $\Delta_\mathcal{C}$, which is a natural geometric object associated to the modes. It is only at the local level of modes that we have formal models. Thus $S$ is approximated by local state spaces $S_X$ corresponding with the modes $X\in\mathcal{C}$ of the system. The idealisation of $\phi$  is replaced in each mode $X$ by a precisely defined function $\phi_X$ on the local state space $S_X$. The values of $\phi_X$ will dictate when the mode must be changed from $X$ to another mode. The monitoring data and control signals for communicating with the environment are measurements specific to each mode, and we use the idea of physical oracle to do this. 

The system \textit{is} the collection of the modes, so we must bring together and integrate the local structures of the modes $X$ and their individual components, such as  $S_X$ and $\mathcal{A}_X$. Here, again, the abstract simplicial complex is used: by viewing the abstract simplicial complex as a category the state and data type components can be brought together by presheaves.  Thus, our model of the architecture of the system has the basic mathematical form of a complicated presheaf.

The evaluation of states relative to modes and the protocols for mode transitions can be quantified and visualised geometrically. Yet again, the abstract simplicial complex is used. From abstract simplicial complexes can be derived concrete \textit{simplicial complexes} in real space  $\mathbb{R}^{n}$. These are  geometric data types and well known in topology: an n-dimensional simplicial complex is essentially a convex hull of $n + 1$ distinct points in a vector space \cite{Span}. Thus, for an abstract simplicial complex $\mathcal{C}$, we can construct a unique geometric simplicial object  $\Delta_\mathcal{C}$ in $\mathbb{R}^{n}$. Note that the geometric and numerical structure $\Delta_\mathcal{C}$ \textit{represents the whole system as it is constructed from the modes}. 

Modes are visualised by the simplices of $\Delta_\mathcal{C}$ which is a data type that can help make a judgement about the data/evidence for contested choices or cooperation between modes. The behaviour of a system is evaluated by mapping its local states into one or more simplices. There is a computable approximation $\phi_X:S_X\to \Delta_\mathcal{C}$ that provides a measure of the extent to which the behaviour is acceptable for several modes.  This is the means to recover the global picture the system, and decide when modes need to be changed. 

The main examples we have in mind are physical systems. But the methods seem to have wider scope and may usefully apply to human-centred systems. For example, classifications in security, privacy, trust and other concepts could be thought of as modes with different restrictions on access (see subsection \ref{scope}).

Finally, let us note that modes and states are widely used terms in computer science and do have several informal or implicit definitions. Commonly they lack a consensus even in particular areas, such as human-computer interaction etc. In software development, the terms appear with different meanings in software design tools even though there is a standard (ISO/IEC/IEEE 24765). An interesting and useful study of the terminology for software tools is Baduel, Bruel, Ober and Doba \cite{Badueletal2018}, which cross-references a number of interpretations of state and modes and draws out some commonalities. Our formulation is consistent with their recommended attributes.

The structure of the paper is as follows. 

In Section \ref{Decision_making}, using simple examples, we give an informal introduction to our concept of modes and of modelling the data involved in making mode transitions. In Section \ref{Covers_and_simplicial_complexes}, we summarise for the convenience of the reader the mathematical ideas about simplicial complexes we need to formalise modes and mode transitions. Then, in the next three sections, we give a full formal account of our model of modes and mode transitions using the mathematics of Section \ref{Covers_and_simplicial_complexes}. In Section \ref{autonomous_racing_cars}, we give a fully formalised case study by applying the framework to autonomous racing cars. In Section \ref{concluding_remarks} we discuss the scope of the modelling technique and its possible technical development.


\section{Aperitif: What is a mode?} \label{Decision_making}

We begin with some examples to illustrate our initial ideas about modelling modes and mode transitions, and give a general informal characterisation.

\subsection{Examples}

\subsubsection{Control: Landing on Mars} \label{marsbay}
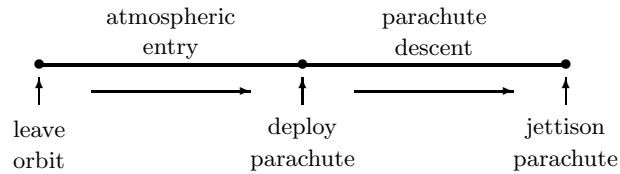
\begin{figure}
\begin{center}
\unitlength 0.7 mm
\begin{picture}(120,35)(10,30)
\linethickness{0.3mm}
\put(20,50){\line(1,0){100}}
\linethickness{0.15mm}
\put(20,42.5){\line(0,1){5}}
\put(20,47.5){\vector(0,1){0.12}}
\linethickness{0.15mm}
\put(70,42.5){\line(0,1){5}}
\put(70,47.5){\vector(0,1){0.12}}
\linethickness{0.15mm}
\put(120,42.5){\line(0,1){5}}
\put(120,47.5){\vector(0,1){0.12}}
\linethickness{0.15mm}
\put(30,45){\line(1,0){30}}
\put(60,45){\vector(1,0){0.12}}
\linethickness{0.15mm}
\put(80,45){\line(1,0){30}}
\put(110,45){\vector(1,0){0.12}}
\put(120,38){\makebox(0,0)[cc]{jettison}}

\put(70,38){\makebox(0,0)[cc]{deploy}}

\put(20,38){\makebox(0,0)[cc]{leave}}

\put(20,32){\makebox(0,0)[cc]{orbit}}

\put(70,32){\makebox(0,0)[cc]{parachute}}

\put(120,32){\makebox(0,0)[cc]{parachute}}

\put(45,59){\makebox(0,0)[cc]{atmospheric}}

\put(45,53){\makebox(0,0)[cc]{entry}}

\put(95,59){\makebox(0,0)[cc]{parachute}}

\put(95,53.5){\makebox(0,0)[cc]{descent}}

\put(120,50){\makebox(0,0)[cc]{$\bullet$}}

\put(70,50){\makebox(0,0)[cc]{$\bullet$}}

\put(20,50){\makebox(0,0)[cc]{$\bullet$}}

\end{picture}
\caption{Visualising a mode transition for a Mars probe}
\label{modpic1}
\end{center}
\end{figure}

A Mars lander goes through the (simplified) five \textit{modes} of: firing engines to leave orbit; atmospheric entry using a heat shield; deploying a parachute; parachute descent; and jettisoning the parachute. As the wreckage of the Mars Polar Lander and the Schiaparelli lander attest \cite{Schiaparelli,Polar}, the important thing is timing. These modes are different, three of them are distinct actions, and two are experiences of significant duration. In Figure~\ref{modpic1}, we show the short duration modes as points, and the longer duration modes as lines connecting the points. Thus, we might visualise the changing of the state of the system as a point travelling from one end of the interval to the other. The picture is like a progress bar.
The continuous progression contributes to the transparency of the process by reassuring any passengers that the procedure for the descent is on schedule.

Note, a standard picture of the system as a graph would look very different -- there would be five vertices linked by arrows, and the state of the system would hop discontinuously from vertex to vertex. The graph loses the visual representation of the continuous progression from one state to another.

\subsubsection{Alert: Avoiding collisions}\label{acol}
Consider some modes that can be found in driving a car. A driver continually monitors for situations which might cause a problem, such as children playing with a ball by the side of the road. 
If the situation changes, such as the ball being kicked across the road, then the driver may anticipate a problem and slow down. 
If a child were to run across the road after the ball, then the driver's priority becomes avoiding a collision. There are obvious options to weigh: 

Brake: will the car stop in time? 

Steer one way: will the car hit the group at the roadside?

Steer the other way: will the car hit the child or an oncoming car? 

\noindent Both braking and steering left, or right, are also options. 

\begin{figure}
\begin{center}
 \unitlength 0.45 mm
\begin{picture}(130,51)(10,27)
\linethickness{0.3mm}
\multiput(70,50)(0.24,0.12){167}{\line(1,0){0.24}}
\linethickness{0.3mm}
\linethickness{0.3mm}
\multiput(70,50)(0.24,-0.12){167}{\line(1,0){0.24}}
\linethickness{0.2mm}
\put(70,50){\line(1,0){55}} 
\linethickness{0.2mm}
\linethickness{0.3mm}
\multiput(110,70)(0.12,-0.16){125}{\line(0,-1){0.16}}
\linethickness{0.3mm}
\multiput(110,30)(0.12,0.16){125}{\line(0,1){0.16}}

\linethickness{0.1mm}
\linethickness{0.1mm}
\multiput(105,37)(0.14,0.12){83}{\line(1,0){0.18}}
\linethickness{0.1mm}
\multiput(109,62)(0.12,-0.13){67}{\line(0,-1){0.13}}
\linethickness{0.1mm}
\multiput(104,60)(0.12,-0.13){42}{\line(1,0){0.12}}
\linethickness{0.1mm}
\multiput(100,40)(0.14,0.12){50}{\line(1,0){0.14}}

\linethickness{0.3mm}
\multiput(89,57)(0.24,0.12){42}{\line(1,0){0.24}}
\put(99,62){\vector(2,1){0.12}}
\linethickness{0.3mm}
\multiput(89,43)(0.24,-0.12){42}{\line(1,0){0.24}}
\put(99,38){\vector(2,-1){0.12}}
\linethickness{0.3mm}
\put(90,52){\line(1,0){10}}
\put(100,52){\vector(1,0){0.12}}
\linethickness{0.3mm}

\put(20,50){\line(1,0){50}}
%

\put(65,57.5){\makebox(0,0)[cc]{collision}}

\put(65,43){\makebox(0,0)[cc]{warning}}

\put(21,57.5){\makebox(0,0)[cc]{normal}}

\put(21,43){\makebox(0,0)[cc]{driving}}

\put(130,70){\makebox(0,0)[cc]{steer left}}

\put(130,30){\makebox(0,0)[cc]{steer right}}

\put(138,51){\makebox(0,0)[cc]{brake}}


\put(170,54){\makebox(0,0)[cc]{avoiding a}}
\put(170,46){\makebox(0,0)[cc]{collision}}

\end{picture}
\caption{Visualising a mode transition for a motor vehicle}
\label{modpic2}
\end{center}
\end{figure}
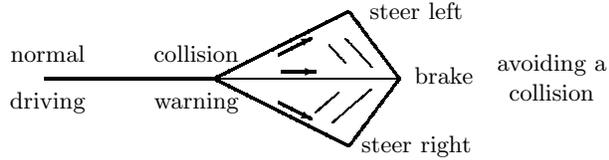

We can picture the decisions to be made in Figure~\ref{modpic2}. As the driver's unease increases, the state of the system moves from normal driving along the edge to collision warning, and the driver may reduce speed. As the child starts running after the ball, we move through collision warning to actively considering the best option for collision avoidance. The state here are two triangles, depending on whether the safest course would be to steer right or left (we cannot do both). Suppose that left is indicated and so we are in the upper triangle. 

A state evaluated in the triangle means that, given the dangers of taking unnecessary avoiding action, we are merely in a state of readiness and information gathering.  If the state is near the vertex \textit{collision warning} then there is little to suggest a mode transition is necessary, where as states further away is evidence that some action is required. A state on the edge between \textit{steer left} and \textit{brake} means that avoiding measures will now be taken.  The measure or combination of the two actions is given by the position along that edge. Note that the joint mode combining both actions may well be much more complicated than implementing the separate actions.

The different possibilities for the action taken mean that we have a mode for avoiding an object that is naturally 2-dimensional. The planning being taken by the autopilot is a continuous change of state in the diagram. 

Note, the standard graph would have 13 vertices, lots of arrows and discontinuous evolution -- the idea of continuous evolution means that being nearby in the environment and in the picture hopefully coincide, so the resulting picture is easier to interpret. 

\subsection{Intuitions about evaluating states and simplical complexes}
Our simple illustrations here introduce a sense of what we are aiming at: (i) a general notion of mode and mode transition; (ii) a way of evaluating, calibrating or interpreting the environment data; and (iii) a way of visualising the evaluation of the states of modes geometrically. 

The raw data we have available about system behaviour are represented by the states of the system. For a real world physical system, commonly there is some space of states based upon physical measurements that are sampled at discrete time points and are approximate.

When we specify the modes of a system then what we need to know about the system is specific to the mode; thus, each mode will have its \text{own} measurable data and state space. Thus, mode transition requires a means of relating the data available to different  modes. These complications will reveal themselves as we build the model. 

The geometric object of choice is a simplical complex which is a higher dimensional analogue of a graph, and is made of $n$-dimensional components, called $n$-simplices for integer $n\ge 0$. For a graph, a 0-simplex is a vertex, and a 1-simplex is a line between two 0-simplices. In general we then have 2-simplices (filled in triangles) between three 1-simplices (edges) which form the boundary of the triangle. Next we have 3-simplices (filled in tetrahedra) between four 2-simplices which form the faces of the tetrahedron, etc. 

\begin{figure}
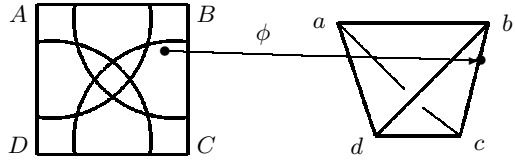

\begin{center}

\unitlength 0.5 mm


\caption{An assessment map $\phi$ and a tetrahedron for a system with 4 modes which all intersect}
\label{assess}
\end{center}
\end{figure}

Thus, from the illustrations, we have a space of measurements, and an {\em assessment map} $\phi$ that classifies or calibrates the states by mapping into a simplicial complex modelling the modes. In Figure~\ref{assess} we have supposed that the space of all measurements is a square, and that the regions associated to the modes are quarter disks centred on the four vertices $A, B, C, D$. 
The highlighted point in the square is in both the regions centred on $B$ and $C$, and  is associated to modes $b$ and $c$ but not to modes $a$ or $d$, and is mapped to a point on the line between $b$ and $c$. 
(Of course, this example looks strange as our space of decisions is higher dimensional than the space of evidence, but in general the opposite will be very much the case.)

\subsection{Principles for modes}\label{mode}
 
A mode is a high-level categorisation of the operation and behaviour of a system.   A mode specifies data about the actual operation of a system together with objectives or tasks for the system in that mode. Typically, the modes are governed by independent models and algorithms operating on their own data, corresponding to different situations. For example, the basis of a classification into modes is determined by the ranges of applicability of certain algorithms on a data types that are a control mechanism of a physical system. 
 
What are the ideas that characterise a system made of modes? 

\medskip\noindent
\textbf{Completeness}. A set of modes for a system is a classification of the operation or behaviour of a system. At any time, a system can be in one, or more, modes. 

\medskip\noindent
\textbf{Localisation}. Each mode for a system chooses and collects its own data to monitor its behaviour and environment. This monitoring data determines the mode's state space and data type for computations.

\medskip\noindent
\textbf{Combination}. When a system is in a number of modes then that situation itself constitutes a mode of the system. 

\medskip\noindent
\textbf{Component}. A set of modes for a system consists of a (i) a set of \textit{basic modes} and (ii) \textit{joint modes} made by combining other modes. 

\medskip\noindent
These are the most important properties of our concept of modes:  if a system is transitioning from basic mode $\alpha$ to basic mode $\beta$  then there is an intermediate or joint mode $\alpha, \beta$ which represents the situation that although the system is in mode $\alpha$ it is aware that it will need to change to mode $\beta$; informally, in symbols, at least three modes are involved: 
$$\{ \alpha \} \to  \{ \alpha, \beta \} \to  \{ \beta \}.$$
The transition could be more complicated.  In transitioning from basic mode $\alpha$ to basic mode $\beta$, one can imagine that $\alpha$ has to make a choice from a number of modes $\gamma_1, \ldots \gamma_k$,  as well as $\beta$; informally, in symbols, four modes are involved: 
$$\{ \alpha \} \to  \{ \alpha, \gamma_1, \ldots \gamma_k, \beta \} \to \{ \alpha, \beta \}  \to  \{ \beta \}.$$

\noindent Fundamental questions arise in deciding 

(i) if and when a system should change from one mode to some other mode, and 

(ii) which new mode should chosen.

\noindent As one mode seems less fit for purpose so other modes may become more relevant. However, as data is inexact or incomplete or faulty, the situations are not always easy to recognise. The change from one mode to some other mode is then an inexact process involving decisions by people or algorithms. 

\smallskip\noindent
\textbf{Quantification}. If a state of the system is meaningful for a number of modes then the relevance or suitability of these modes must be quantified and evaluated.

\smallskip\noindent
\textbf{Thresholds}. The transition out of one mode into one other is governed by the results of the quantification and calibration. The decision to move to a new mode may be specified by numerical thresholds.

\smallskip\noindent
To these may be added postulates about problem situations that require modes to handle exceptional behaviour.

\smallskip\noindent
\textbf{Visualisations}. The quantification of mode states and use of thresholds need to be visualised to suggest instrumentation for modes and transitions.


\section{Abstract and geometric simplicial complexes}\label{Covers_and_simplicial_complexes}

The informal ideas of Section \ref{Decision_making} will be modelled using a series of mathematical concepts that we will now summarise.

\subsection{Abstract simplicial complexes}\label{abstract_simplicial_complexes}

We imagine the behaviour of a system is modelled by a collection of state spaces that are determined by observations of the system in different modes and which cover all likely eventualities. Thus, the {\textit{union}} of the collection is expected to be a \textit{cover} of a global state space for the system. 
 
Let $S$ be a set and take a finite collection of subsets $U_\alpha \subset S$ with indices $\alpha\in\mathcal{M}$ such that $\cup_{\alpha\in X}U_\alpha=S$; this is a \textit{finite cover} of $S$.  Define the \textit{nerve} $\mathcal{C}$ of the cover $\{ U_\alpha : \alpha\in\mathcal{M} \}$ which contains all sets of indexes whose subsets in $S$ overlap:
\[
\mathcal{C} = \big\{ X\subset \mathcal{M} : \cap_{\alpha\in X} U_\alpha\neq\emptyset\big\}\ ,
\]

\begin{defin} \label{absc}
An {\em abstract simplicial complex} $(\mathcal{M},\mathcal{C})$ consists of a collection $\mathcal{C}$ of finite 
subsets  of a set $\mathcal{M}$ with the property that if $Y\subset X$ and $X \in \mathcal{C} $ then $Y\in \mathcal{C} $. An element of 
$X\in \mathcal{C}$ can be called a {\em simplex}, and $Y\subset X$ a {\em sub-simplex} of $X$. 
\end{defin}

\begin{propos} \label{absc56}
The \textit{nerve} $\mathcal{C}$ of the cover $\{ U_\alpha : \alpha\in\mathcal{M} \}$
is an abstract simplicial complex.
\end{propos}
\vskip-0.1in
\noindent\textbf{Proof:}\quad For $X\in \mathcal{C}$ we have $\cap_{\alpha\in X} U_\alpha\neq\emptyset$. Now for $Y\subset X$ we also have $\cap_{\alpha\in Y} U_\alpha\neq\emptyset$. \quad$\square$

%
%

\smallskip
To compare abstract simplicial complexes we use:

\begin{defin} \label{absmaps}
A map of abstract simplical complexes $\Psi:(\mathcal{M},\mathcal{C})\to (\mathcal{M}',\mathcal{C}')$ is a function $\Psi:\mathcal{M}\to \mathcal{M}'$ so that on subsets
if $X\in \mathcal{C}$ then $\Psi X\in \mathcal{C}'$. 
\end{defin}

The refinement of a cover yields a simple example. 
Note that we demand a map on indices for this definition. Refinement of covers is a method of giving a hierarchy, where we have refinements of refinements, and so on. 

\begin{defin} \label{]refine}
A cover $\{W_k:k\in\mathcal{N}\}$ of $S$ is a {\em subcover} of the cover $\{U_\alpha:\alpha\in\mathcal{M}\}$ if there is a map 
$\Psi:\mathcal{N} \to \mathcal{M}$ so that $W_k\subset U_{\Psi(k)}$. 
\end{defin}
This extends to a map of abstract simplical complexes on the nerves of the covers.

\subsection{Realisation of abstract simplicial complexes in $\mathbb{R}^{n}$}\label{real_simplicial_complexes}

\begin{defin} \label{realsc}
A concrete {\em simplicial complex} is a collection of simplices in some vector space so that the face of any simplex in the collection is also in the collection, and the intersection of any two simplices is a face of both of them.
\end{defin}

\begin{propos}\label{Real_representation}
To every abstract simplicial complex $(\mathcal{M},\mathcal{C})$ (as in Definition~\ref{absc}) is associated its realisation $\Delta_\mathcal{C} \subset \mathbb{R}^\mathcal{M}$, as a simplical complex. 
\end{propos}
\vskip-0.1in
\noindent\textbf{Proof:}\quad 
Form a vector space $\mathbb{R}^\mathcal{M}$ with basis $e_\alpha$ for $\alpha\in \mathcal{M}$. The simplex spanned by $X\in \mathcal{C}$ is
\[
\Delta_X=\Big\{ \sum_{x\in X} \lambda_x  \, e_x : 
\lambda_x\in[0,1],\  \sum_{x\in X} \lambda_x=1\Big\}. \qquad\square
\]

\begin{propos}
A map of abstract simplical complexes $\Psi:(\mathcal{M},\mathcal{C})\to (\mathcal{M}',\mathcal{C}')$ can be expended to their realisations
as $\Delta_\Psi:\Delta_\mathcal{C} \to \Delta_\mathcal{C'} $ by defining
\[
\Delta_\Psi\Big(  \sum_{x\in X} \lambda_x  \, e_x  \Big) = \sum_{x\in X} \lambda_x  \, e_{\Psi(x)}.
\]
\end{propos}

\medskip
The simplex $\Delta_X$ is a $(|X|-1)$-simplex where $|X|$ is the size of $X$, and if $Y\subset X$ then $\Delta_Y$ is a face of $\Delta_X$. 
Then $\Delta_X\cap \Delta_Z=\Delta_{X\cap Z}$ and Definition~\ref{realsc} is seen to be satisfied.

Having created the geometric simplex we are now able to use the categorisation of the data in a set $S$ according to a cover $\{ U_\alpha : \alpha\in\mathcal{M} \}$, associated with an abstract simplical complex $\mathcal{C}$, to create a geometric visualisation in $\Delta_\mathcal{C}$. To do this we simply need an appropriate map from  $S$ to $\mathcal{C}$.

\begin{figure}
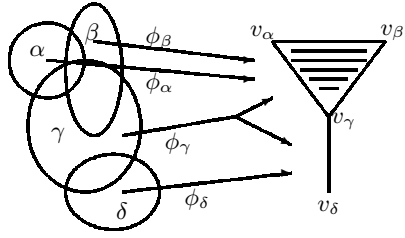

\begin{center}
\unitlength 0.5 mm


\caption{A partition of unity as a refinement of a cover by four sets}
\label{part4}
\end{center}
\end{figure}

\begin{defin} \label{realpu}
A {\em partition of unity} for the cover $U_\alpha\subset S$ is a function $\phi_\alpha:S\to[0,1]$ for every $\alpha\in\mathcal{M}$ such that 

(1)\quad
if $\phi_\alpha(s)\neq 0$ then $s\in U_\alpha$;

(2)\quad $\sum_{\alpha\in\mathcal{M}}\phi_\alpha(s)=1$ for all $s\in S$.

\noindent We then have a function $\phi:S\to \Delta_{\mathcal{C}}$ given by
$$\phi(s)= \sum_{\alpha\in\mathcal{M}} \phi_\alpha(s)\,e_\alpha.$$

\end{defin}
Figure~\ref{part4} visualises a simplicial complex and partition of unity for a cover by four sets.
Note that the triangle in  Figure~\ref{part4} is shaded to form a 2-simplex precisely because $U_\alpha\cap U_\beta\cap U_\gamma$ is not empty.  For specific circumstances we can impose extra conditions on $\phi$, e.g.,\ continuity or computability.

\subsection{Product simplical complexes and twofold covers} \label{prodtwo}

Given abstract simplicial complexes $\mathcal{C}\subset P(\mathcal{M})$ and $\mathcal{D}\subset P(\mathcal{N})$, we have a product complex $\mathcal{C} \times \mathcal{D}$
with vertices $(c,d)\in \mathcal{M}\times\mathcal{N}$ and with simplices
\begin{align} \label{prodcomp}
\big\{  \{ (c_1,d_1),\dots ,(c_k,d_k) \} \in P(\mathcal{M}\times\mathcal{N}) :
\{c_1,\dots, c_k\}\in \mathcal{C}\  \& \ \{d_1,\dots, d_k\}\in \mathcal{D}  \big\}.
\end{align}
There are maps of abstract simplical complexes $\pi_1:\mathcal{C} \times \mathcal{D} \to \mathcal{C}$ and  $\pi_2:\mathcal{C} \times \mathcal{D} \to \mathcal{D}$ given by
\begin{align*}
\pi_1\{ (c_1,d_1),\dots ,(c_k,d_k) \} &=\{ c_1,\dots ,c_k \}  \cr
\pi_2\{ (c_1,d_1),\dots ,(c_k,d_k) \} &=\{ d_1,\dots ,d_k \} \ .
\end{align*}
Given two covers $U_\alpha$ (for $\alpha\in A$) and $V_i$ (for $i\in I$) of $S$, we can take the intersection cover $U_\alpha\cap V_i$, and the nerve of the intersection cover $\mathcal{UV}$ is a subcomplex of the product of the nerves $\mathcal{U}$ and $\mathcal{V}$ of the individual covers, and maps to each of the nerves of the original covers. We call this 
nerve of the intersection complex the nerve of the twofold cover, and this can be generalised to $n$-fold covers.

\begin{propos}
Given a partition of unity $\phi_{\alpha,i}$ for the cover $U_\alpha\cap V_i$ and elements $e_{\alpha,i}$ of a vector space, we have a map of realisations
\begin{eqnarray*}
\UseComputerModernTips
\xymatrix{
\Delta_\mathcal{U} & \Delta_\mathcal{UV}  \ar[r]_{\pi_\mathcal{V}}   \ar[l]^{\pi_\mathcal{U}}  &   \Delta_\mathcal{V}    \\ 
	&     S  \ar[u]_{  \chi }  \ar[ur]_{  \psi }    \ar[ul]^{  \phi }    & }
\end{eqnarray*}
where the maps $\pi_\mathcal{U}$ and $\pi_\mathcal{V}$ and the single cover partitions of unity are given by
\begin{align*}
&{\pi_\mathcal{U}} \Big(\sum_{\alpha,i}  \lambda_{\alpha,i}\, e_{\alpha,i}\Big) = \sum_\alpha \Big(
\sum_i \lambda_{\alpha,i}\Big) e_\alpha\ ,\quad {\pi_\mathcal{V}} \Big(\sum_{\alpha,i}  \lambda_{\alpha,i}\, e_{\alpha,i}\Big) = \sum_i \Big(
\sum_\alpha \lambda_{\alpha,i}\Big) e_i \cr
& \phi_\alpha=\sum_i \chi_{\alpha,i}\ ,\quad \psi_i=\sum_\alpha \chi_{\alpha,i}\ .
\end{align*}
An obvious choice of the twofold partition of unity $\chi_{\alpha,i}$ in terms of partitions $\phi_\alpha$ and $\psi_i$ for the single covers would be the product $\chi_{\alpha i}=\phi_\alpha\, \psi_i$.
\end{propos}
\noindent\textbf{Proof:}\quad 
Given that the $ \chi_{\alpha,i}$ are a partition of unity for the cover $U_\alpha\cap V_i$ it is automatic that  the $\phi_\alpha$ and $\psi_i$ in the statement are positive, and they trivially sum to 1. If $\phi_\alpha(x)>0$ then we must have
$ \chi_{\alpha,i}(x)>0$ for some $i$, and as $U_\alpha\cap V_i\subset U_\alpha$ we see that $ x\in U_\alpha$. Similarly for the $\psi_i$. 
Finally note that for the special case of the product partition of unity, if $\phi_\alpha(x) \psi_i(x)>0$ we need both $x\in U_\alpha$ and $x\in V_i$, and that the sum over $\alpha,i$ is 1. \hfill $\square$

\medskip 
Note that our description defines the categorical product, there is a smaller product complex which requires choosing an order $\le$ on both
$\mathcal{M}$ and $\mathcal{N}$, and then insisting that the $c_i$ and $d_i$ in (\ref{prodcomp}) are in increasing order. 
We do not use this since the indices in the pairings $U_\alpha\cap V_i$ have a real meaning and no sensible order, as discussed in 
Section~\ref{model}.

\subsection{Categories, functors and presheaves} \label{sectcat}

\begin{defin} \label{catdef}
A {\em category} consists of a
\begin{enumerate}[topsep=2pt]
\itemsep0em 

\item[{\rm (1)}] collection of {\em objects}\index{object of category} $X,Y,V,W,\dots$.

\item[{\rm (2)}]  specification of a set $\mathrm{Mor}(X,Y)$ of {\em morphisms}\index{morphism} for any objects $X,Y$.

\item[{\rm (3)}]  associative composition operation $\circ:\mathrm{Mor}(Y,Z)\times\mathrm{Mor}(X,Y)\to
\mathrm{Mor}(X,Z)$.

\item[{\rm (4)}] Every set $\mathrm{Mor}(X,X)$ contains an identity element
$\id_X$ such that $\theta\circ\id_X=\theta$ and $\id_X\circ\theta=\theta$ for any 
morphism $\theta$ for which $\circ$ is defined.
\end{enumerate}
\end{defin}

\begin{propos}
From an abstract simplical complex $(\mathcal{M},\mathcal{C})$ we can form a category (which we just call $\mathcal{C}$) which has objects $X\in\mathcal{C}$. For morphisms,
$\mathrm{Mor}(Y,X)$ consists of either one morphism, the arrow $Y\to X$, if $Y\subset X$, or no morphisms if $Y\not\subset X$. 
\end{propos}

Recall that the appropriate idea of a map between categories is a functor.

\begin{defin}
A {\em(covariant) functor} $F:\mathcal{C}\to \mathcal{D}$ between categories
specifies an object $F(X)\in \mathcal{D}$ for every object $X\in\mathcal{C}$, 
and a morphism $F(\theta):F(X)\to F(Y)$ for every morphism
$\theta:X\to Y$, such that $F(\id_X)=\id_{F(X)}$ and
 $F(\theta\circ\psi)=F(\theta)\circ F(\psi)$
for morphisms.
\end{defin}

The following definition corresponds to the topological idea of a presheaf\footnote{These aremost commonly used with the category of abelian groups, except that we consider a fixed cover, not a limit of covers}.

\begin{defin} \label{defmor}
For a category $\mathcal{D}$, a $\mathcal{D}$-valued {\em presheaf} on $\mathcal{C}$ is a functor from $\mathcal{C}$ to 
$\mathcal{D}$. 
\end{defin}
This means that for every $X\in \mathcal{C}$ we have an object
$D_X\in\mathcal{D}$ and that to every  $Y\subset X\in \mathcal{C}$ we have a morphism ${}_X\mathrm{inc}_Y:D_Y\to D_X$. Further, for all $Z\subset Y\subset X\in \mathcal{C}$
we have ${}_X\mathrm{inc}_Y\circ {}_Y\mathrm{inc}_Z={}_X\mathrm{inc}_Z$. 


\section{Modes and their mathematical model} \label{model} 

In Section \ref{Decision_making}, we described some examples and general characteristics to motivate the idea of modes and their mathematical representation by objects in $\mathbb{R}^{n}$.  In  Section \ref{Covers_and_simplicial_complexes} we summarised the mathematics of abstract and concrete simplicial complexes. Now we build the complete model of modes and modes transitions.

\subsection{Modes and abstract simplicial complexes} \label{realc} 

Imagine a physical system that has a number of modes of operation or behaviour. In simple terms, the modes are determined by distinct 
\smallskip

1. \textit{Mathematical models of the system behaviour}, i.e.,\ which data, equations or algorithms best describe the dynamics in this mode.

\smallskip

2. \textit{Intentions and objectives}, i.e.,\ what should the system be doing, what decisions could or should be made in this mode.
To add to our intuitive examples:

\begin{exam} \label{yips}
{\em{
Objectives of modes can be illustrated by a plane which has several distinct modes of operation that determine its response to the controls. Most obviously, it can start, taxi on the ground, take off, climb, cruise, descend, land, and  stop. To these 8 basic modes there are refinements for within flight we can have different control regimes, e.g., \ an aircraft in turbulence or a stall has radically different handling to normal flight. Continuing our aircraft example, a plane running short of fuel has several objectives to chose from: (a) continue to its destination, (b) divert to a nearer airfield or (c) crash landing. These decisions may subdivide, e.g.,\ into (b1) nearest airport with full  emergency facilities or (b2) any landing strip. Different data and mathematical models and needed for these these modes.
}}
\end{exam}

\smallskip

Imagine that the actual behaviour of the \textit{whole} physical system can be expressed by states and trajectories in a state space $S$. We can begin to formalise the properties of modes highlighted in Section \ref{mode} by a categorisation of states: we suppose there is a cover of the state space $S$ by sets, and the set of states that `belong' to a mode is an element of the cover of the state space.  The intersections of sets in the cover have a natural interpretation: a state belonging to two or more elements of the cover means that the state can be in any of these modes. 

So, for example, a state may be subject to processing by the distinct algorithms associated with each set in the cover. The question arises: Which set of algorithms \textit{should} be in control?

Let us go into more detail: imagine we have a state space $S$ for a system in which there is a time evolution of a state $s(t)\in S$ for time $t$. To identify and specify a set $\mathcal{M}$ of modes of behaviour, we localise the behaviour using a cover of sets $U_\alpha$ with index set $\alpha \in \mathcal{M}$.  A state $s \in S$ of the system is in mode $\alpha$ if $s \in U_\alpha$.  Now, we take the nerve of this cover:
\[
\mathcal{C} = \big\{ X\subset \mathcal{M} : \cap_{\alpha\in X} U_\alpha\neq\emptyset\big\}.\ 
\]
The set $\mathcal{C}$ contains all sets of indexes whose associated subsets in $S$ overlap, e.g., $X=\{\alpha,\beta\}\in \mathcal{C}$ if, and only if, $U_\alpha\cap U_\beta$ is not empty.

Recalling Section~\ref{real_simplicial_complexes},  we can immediately make the euclidean realisation  $\Delta_\mathcal{C}\subset \mathbb{R}^\mathcal{M}$ of the simplicial complex $\mathcal{C}$.  We take a partition of unity
\[ 
\phi:S\to \Delta_\mathcal{C}.
\] 
Then, for a given state $s \in S$ of the system we have 
$$\phi(s)=\sum_{\alpha\in\mathcal{M} } \phi_\alpha(s)\,e_\alpha,$$
where $e_\alpha$ for $\alpha\in\mathcal{M}$ is the basis of 
 $\mathbb{R}^\mathcal{M}$.  If $\phi_\alpha(s)>0$ we know that $s\in U_\alpha$. 


\subsection{Local components of modes} \label{compstr}

The specification of the system is entirely built by bringing together the specifications its modes. Thus, we have only the components that belong to modes and our job is to construct a computational structure embracing all the data and algorithms belonging to the modes. Let us take stock: we have an abstract simplicial complex $\mathcal{C}\subset P(\mathcal{M})$ to label the modes of the system; each $\alpha\in \mathcal{M}$ is called a `basic mode' of the system; and a realisation $\Delta_{\mathcal{C}} \subset \R^\mathcal{M}$ of the abstract simplicial complex.

To each mode $X$ we associate a \textit{package} $D_X$  of \textit{mode components}:

\begin{enumerate}
\itemsep0em 

\item For each mode $X\in \mathcal{C}$, a set $S_X$ of states that defines the data available to the system in mode $X$.
 
\item For each mode $X\in \mathcal{C}$, an evaluation function $\phi_X:S_X\to \Delta_{\mathcal{C}}$ that calibrates the state against the modes.

\item For each mode $X\in \mathcal{C}$, an algorithm $\mathcal{B}_X$ based upon a data type $\mathcal{A}_X$.

\item For any modes $\emptyset\neq Y\subset X\in \mathcal{C}$, partially defined functions ${}_X\mathrm{inc}_Y:S_Y\to S_X$  and ${}_Y\mathrm{proj}_X:S_X\to S_Y$ that relate the data available to the modes $X$ and $Y$.

\end{enumerate}
However, our modelling raises a number of subtle computational points about the interface between our computational system and reality:

\medskip
\noindent \textit{Data.} The numerical data available to a mode $X$ depend upon measurements of the system and environment, sampled at various times. This means essentially that the numerical data that makes up each $S_X$ are approximations. Thus, since measurements are rational numbers, if $S_X$ contains only measurements then $S_X \subset  \Q^k$ for some $k$ depending on the mode.

\medskip
\noindent \textit{Computability.} The components of a mode and their inter-dependencies are assumed to be computable.

\medskip
In our computable approximation to the environment we use a set $S_X$ to contain information about the state of the system in mode $X$ (measurements, approximations, predictions,...) and also about our intent (the orders or instructions for the control system). 

To discuss the interface it is convenient to imagine an idealised global state space $S$ to guide our thinking. Of course, the point of modes is that we do not have a workable understanding of the entire system but only of certain modes of operations. Hence, $S$ is an idealisation that is \textit{not} part of the model of the system. If the real system is in state $s\in S$ and $\phi(s)\in\Delta_X$ for some $X\in\mathcal{C}$, then $s\in S$ has a description using data $\tilde s\in S_X$. Our modelling began with a collection of sets that we hypothesise are a cover of $S$. In an ideal case, the state space for mode $X$ would be $S_X = \bigcap_{\alpha\in X} U_\alpha\subset S$. 

Since we have no \textit{direct} access to a global state space $S$, we cannot compute $\phi:S\to  \Delta_{\mathcal{C}}$.  We do have a local computable approximation local to mode $X$: 
\begin{center}
$\phi_X:S_X\to \Delta_{\mathcal{C}}$ computes $\phi_X(\tilde s)$. 
\end{center}

In mode $X$ we have an algebra $\mathcal{A}_X$ of operations and sets, including the local state space $S_X$ and the function $\phi_X$. Using this algebra we write an algorithm $\mathcal{B}_X$ to control the system for as long as mode $X$ is considered a suitable mode to be in control. 

Suppose that the real system follows a path $s(t)\in S$ over time $t$. Then while in mode $X$ we have a path in data-space $\tilde s(t)\in S_X$, which is calculated by the package $D_X$ using various measurements.  In time, the path $s(t)$ will likely leave the subset of $S$ corresponding to mode $X$ and enter that of at least one other mode $Y$. To implement a change of mode, the algorithm will invoke functions  ${}_X\mathrm{inc}_Y$ or ${}_Y\mathrm{proj}_X$ to change the local state space, where it is 
 convenient to separate the cases where we move to a larger ($X\subset Y$) or a smaller ($Y\subset X$) mode. As these functions will be invoked only when a change of mode is being considered according to the data in $S_X$, in general they are only partial functions from $S_X$ to $S_Y$. 

To summarise the role of key components: in a given mode $X\in\mathcal{C}$, the algorithm $\mathcal{B}_X$ controls the system using the data available in $S_X$ and the functions in the algebra $\mathcal{A}_X$. The function $\phi_X$ gives information about which mode we are in, and the partial functions ${}_X\mathrm{inc}_Y$ or ${}_Y\mathrm{proj}_X$ implement the change of mode.

\smallskip\noindent
\textbf{Computation in mode $X$ and the package $D_X$:}
For each mode $X\in \mathcal{C}$ we have the following package $D_X$: 

\begin{enumerate}
\itemsep0em 
   \item A {\em many sorted algebra} $\mathcal{A}_X$ which
    contains, among other things
   \begin{itemize}
   \itemsep0em 
     \item \textbf{Types}  including $S_X$, $\mathcal{C}$, $\Delta_\mathcal{C}$, $check$ \dots
     \item \textbf{Functions} including $\phi_X: S_X\to \Delta_\mathcal{C}$, \dots
          \item \textbf{Relations}
   \end{itemize}
      \item Extensions of the many sorted algebra $\mathcal{A}_X$ which take the form of constants and functions, namely
   \begin{itemize}
   \itemsep0em 
          \item \textbf{Oracles} -- for i/o data for a mode, explained in Section~\ref{realpr}
           \item \textbf{Transfers} -- to exchange data between modes, explained in Section~\ref{transfer}
           \item \textbf{Thresholds} -- to decide a change of modes, explained in Section~\ref{funjup}
              \end{itemize}
   \item An algorithm $\mathcal{B}_X$ using only types, functions, relations, oracles and transfers etc. from the algebra $\mathcal{A}_X$. 
\end{enumerate}

A possible idealised form or template for the algorithm $\mathcal{B}_X$ for mode $X$ is this:
\newline
\newline
\noindent
\noindent\fbox{
\parbox{\textwidth}{%

 {\fontfamily{pcr}\selectfont  \textbf{declaration }}  $\mathit{state}: S_X$;  state component variables; auxiliary variables for calculations; \\
$\mathit{checkTransfer} : \mathrm{check}$

 {\fontfamily{pcr}\selectfont  \textbf{input}} $\mathit{state}$  
 
 {\fontfamily{pcr}\selectfont  \textbf{loop }}

\quad request and receive monitoring data from the environment

\quad  update $\mathit{state}$ with new monitoring data

\quad evaluate $\phi_X(\mathit{state})$ 

\quad compute actions

\quad send instructions to environment

\quad compute `best' mode $Y$ 

\quad  {\fontfamily{pcr}\selectfont  \textbf{if}} $Trigger$ 
{\fontfamily{pcr}\selectfont  \textbf{then}} 
 $ \mathit{checkTransfer}:=\mathrm{tran}(   \mathit{state}, Y    )$

%

\quad  {\fontfamily{pcr}\selectfont  \textbf{if}} $\mathit{checkTransfer}=\mathrm{NotOK}$ 
{\fontfamily{pcr}\selectfont  \textbf{then}} 
exception

 {\fontfamily{pcr}\selectfont  \textbf{return }}
}
}
\newline
\newline
A central task of the algorithm is to decide, while in mode $X$, given a state in $S_X$, should we go to new mode $Y$ or not? 

We must now model the mechanisms involved. The oracles that take care of the environment are easy to formulate, which we do next.  The semantics of a transfer between modes is more complicated, as are the evaluation of the state with respect to modes and the use of thresholds to trigger a transition. These latter processes occupy Sections \ref{transfer} and \ref{thresholds}.

\subsection{Interface with the environment: oracles}\label{realpr}
\begin{center}
``Reality is merely an illusion, albeit a very persistent one"  --  Albert Einstein (Attributed). 
\end{center}

Here we look at the interface between our model and the environment and the procedure by which the algorithm interacts with external reality.  Turing used a term \textit{oracle} for an interaction of an algorithm with a \textit{unknown} device, and we shall generalise this idea slightly, so that the oracles may be used both as input and output devices. Communication with the environment is by oracle calls in control algorithm of the mode, and are implemented using i/o constructs compatible with the programming language of the algorithm. 

This interface means that formally algorithms are not dependent on the means of gathering monitoring data. We adapt the operations in an algebra to model interactions with the environment. 

\begin{defin} \label{oracledef} 
An \textit{oracle} is a function of the form
\begin{center}
$\mathcal{O}: \mathrm{output}\times \mathrm{environment} \to \mathrm{input}\times \mathrm{environment}$. 
\end{center} 
\noindent In terms of the many sorted algebra, it appears as a function
\begin{center}
$\mathcal{O}: \mathrm{output}\to \mathrm{input}$
\end{center} 
where  $\mathrm{output}$ and $ \mathrm{input}$ are types, and with the caveat that the value of the function is not repeatable: given an input, it may give different values, simply because the state of the environment may have changed. 
\end{defin}
Thus, the environment itself is invisible to the algebra, only its indirect effects on the values returned by the oracles can be noticed. How do these functions work? Here are two examples. The first  simply collects data from the environment for the algorithm and the second issues an instruction to the environment.

\begin{exam} \label{posmes}
Consider a racing car on a track of length $L$. To establish its position $x\in[0,L]$, we take a measurement oracle 
$\mathcal{O}_x: \{0\}\times \mathrm{environment} \to [0,L]\times \mathrm{environment}$. The $ \{0\}$ simply means that the call of the function has a trivial parameter;  the function is called without an argument. Given a state of the system $r\in \mathrm{environment}$, we have $\mathcal{O}_x(0,r)=(x(r),r)$, where we take $x(r)$ to be the position of the car in that state of the system. The value of $r$ has not changed. (In fact the measurement may have changed the system slightly, but here we ignore that.) We also assume that the measurement is instantaneous, if the system was subject to delay we should include a time stamp on the measurement, giving an output $(x,t)\in [0,L]\times T$ where $T$ is the set for time and $t\in T$ is when the observation was made. 
\end{exam}

The second example effects a change in the environment but may or may not return a confirmation message to the algorithm.

\begin{exam} \label{gearset}
Consider a mechanical system with an actuator that rotates a disk.  An instruction to an actuator has the form
$\mathcal{O}_{rot}: \Z\times \mathrm{environment} \to \{0\}\times \mathrm{environment}$. The effect on the environment is $\mathcal{O}_{rot}(n,r)=(0,\mathrm{turn}(r,n))$,  as $\mathrm{turn}(r,n)$ takes the system in state $r$ and turns the disk by $n^\circ$.  However, suppose that the algorithm does need to know that the task was successfully completed. Then we could specify 
$\mathcal{O}_{rot}: \Z\times \mathrm{environment} \to  \{ \mathrm{OK}, \mathrm{NotOK} \}  \times \mathrm{environment}$, where returning OK to the algorithm means that the device doing the turning thinks it was successful, and NotOK indicates a problem.
\end{exam}


\section{Mode transitions: Transferring control to another mode}\label{transfer}

We have considered the role of the computational packages $D_X$ which are indexed by the modes $X\in \mathcal{C}$. Now we consider the morphisms between the components, which are involved in the transfer of control from one mode to another. In this section, we concentrate on the state space $S_X$ and the information available to the mode $X$.
As we move from mode $X$ to mode $Y$,  how does $S_X$ change and relate to $S_Y$? There are two cases.

\subsection{Moving to a superset: $S_Y\to S_X$ for $Y\subset X$}\label{states_presheaves}

Consider an example in which there is a change from a simpler to a more complicated situation, in that more factors need to be taken into account.

\begin{exam} \label{brake3}
We refer back to Section~\ref{acol}, and use $SL$ for `steer left' and $B$ for `brake'. For the modes $\{SL\}$, $\{B\}$ and  $\{SL,B\}$  we have the corresponding state spaces, $S_{SL}$, $S_B$ and $S_{  \{SL,B\} }$. Now suppose that the car is in braking mode $B$, but that sensor data indicate that to avoid a collision it is also necessary to turn left. Now the car must rapidly fill in information about what is to its left, among other things necessary to safely steer left. In terms of our data structure, $S_B$ is copied into $S_{  \{SL,B\} }$ -- there are no deletions as information relevant to braking is relevant to the joint operation. However $S_{  \{SL,B\} }$ will need some placeholders saying `fill in with sensor observations as soon as possible'. 
\end{exam}

There are three general ideas that arise in such cases. 

\begin{enumerate}
\itemsep0em 
\item \textit{Data}. Initially, in moving to a more complicated situation we do not delete data which we already have. This may be done subsequently, but at first sight we do not know how much of our current information is relevant. 

\item \textit{Partiality}. Moving from mode $\alpha$ to $\{\alpha,\beta\}$ may not make sense for all states of the system $S_\alpha$, so the map 
$S_\alpha \to S_{\{\alpha,\beta\}}$
on the state space is likely to be only a partial map.

\item \textit{Timing}. Two moves in quick succession, say from $S_\alpha$ to $S_{\{\alpha,\beta\}}$ and then to $S_{\{\alpha,\beta,\gamma\}}$
(with no time to gather or calculate more information in between) would be the same as moving directly from $S_\alpha$ to $S_{\{\alpha,\beta,\gamma\}}$. 
\end{enumerate}

\noindent These ideas are contained in the following mathematical principle:

\begin{principle} \label{prin1}
For all
 $\emptyset\neq Y\subset X\in \mathcal{C}$ we have an injective partial map ${}_X\mathrm{inc}_Y:S_Y\to S_X$. For all
  $\emptyset\neq Y\subset X\subset Z\in \mathcal{C}$ we have ${}_Z\mathrm{inc}_X\circ {}_X\mathrm{inc}_Y={}_Z\mathrm{inc}_Y$ (the functorial property for $\mathrm{inc}$). For completeness, set ${}_X\mathrm{inc}_X:S_X\to S_X$ to be the identity.
\end{principle}

Property 2 leads us to consider the category $\mathcal{S}$ of sets and partial maps.
We use Definition~\ref{defmor} to formalise property 3 as a presheaf on the simplical complex $\mathcal{C}$.

\begin{lem} \label{pre6}
For the category $\mathcal{S}$ of sets and partial functions, a
 $\mathcal{S}$-valued simplicial presheaf is a functor $(S,\mathrm{inc})$ from an abstract simplical complex $\mathcal{C}\subset P( \mathcal{M})$, with morphisms being inclusion $Y\subset X$, to the category $\mathcal{S}$. This means that for every $X\in \mathcal{C}$ we have a set
$S_X$ and that to every  $Y\subset X\in \mathcal{C}$ we have a partial map ${}_X\mathrm{inc}_Y:S_Y\to S_X$. Further, for all $Z\subset Y\subset X\in \mathcal{C}$
we have ${}_X\mathrm{inc}_Y\circ {}_Y\mathrm{inc}_Z={}_X\mathrm{inc}_Z$. 
\end{lem}

\subsection{Moving to a subset: $S_X\to S_Y$ for $Y\subset X$}\label{states_presheaves_sub}

Consider an example in which there is a change from a more complicated situation to a simpler one, in that fewer factors need to be taken into account.

\begin{exam} \label{brake9}
Continuing from Example~\ref{brake3}, we can consider a car doing both $SL$ (steer left) and $B$ (brake) to avoid a collision. If the brake is only being touched very lightly, if at all, it would not cause major disruption to move to the single mode $SL$. However, if the brake is being applied heavily when the system tries to move into  the single mode $SL$ the result is likely to be a major panic, and an urgent move back into the joint mode. 
\end{exam}

Again, there are three general ideas that arise such cases. 
\begin{enumerate}
\itemsep0em 
\item \textit{Data.} We may delete data which would be relevant only  for factors which are no longer considered. 

\item \textit{Partiality.} Moving from mode  $\{\alpha,\beta\}$ to $\alpha$ only makes sense if the basic mode $\beta$ is no longer considered relevant, so the map 
$S_{\{\alpha,\beta\}} \to S_\alpha$
on the state space is likely to be only a partial map.

\item \textit{Timing.} Two moves in quick succession, say from $S_{\{\alpha,\beta,\gamma\}}$ to $S_{\{\alpha,\beta\}}$ and then to $S_\alpha$ 
(with no time to gather or calculate more information in between) would be the same as moving directly from $S_{\{\alpha,\beta,\gamma\}}$  to $S_\alpha$. 

\end{enumerate}

\noindent These ideas are contained in the following mathematical principle:

\begin{principle} \label{prin2}
For all
 $\emptyset\neq X\subset Y\in \mathcal{C}$ we have a partial map ${}_X\mathrm{proj}_Y:S_Y\to S_X$. For all
  $\emptyset\neq Z\subset X\subset Y\in \mathcal{C}$ we have ${}_Z\mathrm{proj}_X\circ {}_X\mathrm{proj}_Y={}_Z\mathrm{proj}_Y$ (the functorial property for $\mathrm{proj}$). For completeness set ${}_X\mathrm{proj}_X:S_X\to S_X$ to be the identity.
\end{principle}

\subsection{Combining supersets and subsets, and subsets and supersets}\label{supersub}

The two data properties in \ref{states_presheaves} and \ref{states_presheaves_sub} have an asymmetry where one map is injective (as data must be copied) and the other is not (as deletions can happen).  In the case of the timing properties, two moves in quick succession, say from $S_\alpha$  to $S_{\{\alpha,\beta\}}$ and then back to $S_\alpha$ (with no time to gather or calculate more information in between) would be the same as the identity from $S_\alpha$  to $S_\alpha$ (at least on the domain of the partial map $S_\alpha\to S_{\{\alpha,\beta\}}$). We can use these and other observations to motivate the following principle. 

\begin{principle} \label{prin3}
For all
 $\emptyset\neq X\subset Y\in \mathcal{C}$ we have ${}_X\mathrm{proj}_Y\circ {}_Y\mathrm{inc}_X:S_X\to S_X$ being the identity on the domain of ${}_Y\mathrm{inc}_X:S_X\to S_Y$.
\end{principle}

According to whether we are moving along the arrows in $\mathcal{C}$ or in the opposite direction we 
 distinguish the two partial functions 
\begin{center}
${}_Z\mathrm{inc}{}_X:S_X\to S_Z$ for $X\subset Z$ and ${}_Z\mathrm{proj}{}_X:S_X\to S_Z$ for $Z\subset X$. 
\end{center}
To invoke these while in mode $X$ we shall simply use one `function' 
\begin{center}
$\mathrm{tran}:S_X\times\mathcal{C}\to \mathrm{Check}$
\end{center} 
in the many sorted algebra $\mathcal{A}_X$, which if successful will transfer control from the mode. The set $\mathrm{Check}$ is a set of exceptions which can be thrown if the transfer does not take place. In mode $X$ the call $\mathrm{tran}(s,Z)$ will have the result:
 
 \smallskip\noindent
 a) If $Z=X$ then no transfer will take place and the function will return the value $\mathit{OK}\in \mathrm{Check}$. 
 
 \noindent
 b) If $X\subset Z$ and $X\neq Z$ then (if successful) the algorithm $\mathcal{B}_X$ will terminate and 
 $\mathcal{B}_Z$ will be started with initialisation $\mathit{state}={}_Z\mathrm{inc}{}_X(s)\in S_Z$. 

 \noindent
 c) If $Z\subset X$ and $X\neq Z$ then (if successful) the algorithm $\mathcal{B}_X$ will terminate and 
 $\mathcal{B}_Z$ will be started with initialisation $\mathit{state}={}_Z\mathrm{proj}{}_X(s)\in S_Z$. 
 
 \noindent
 d) If none of $X\subset Z$, $X= Z$, $Z\subset X$ are true then no transfer will take place and the function will return the value $\mathit{notOK}\in \mathrm{Check}$.


\section{Thresholds for invoking a transition}\label{thresholds}

What can trigger a mode transition? To resolve this we look in more detail at simplicial complexes. 

\subsection{Paths on simplicial complexes}  \label{paths6}
We use the notation for the realisation of a simplicial complex $\Delta_\mathcal{C}$ from Section~\ref{real_simplicial_complexes}. Then $\Delta_\mathcal{C}$ is the union of the simplices $\Delta_X$ for $X\in \mathcal{C}$. We define the interior of $\Delta_X$  to be
\begin{align} \label{inter}
\mathrm{int}\Delta_X = \Big\{ \sum_{\alpha\in X} \lambda_\alpha  \, e_\alpha : 
\lambda_\alpha\in(0,1],\  \sum_{\alpha\in X} \lambda_\alpha=1\Big\}
\end{align}
in the same terms as Proposition~\ref{Real_representation}, the difference being that all $\lambda_{\alpha}>0$ in the interior whereas we just have $\lambda_{\alpha}\ge 0$ in the simplex. 

\begin{lem}
For every element $v\in \Delta_\mathcal{C}$ there is a unique $X\in\mathcal{C}$ such that  $v\in \mathrm{int} \Delta_X$. If we write $v=\sum_{\alpha\in \mathcal{M}} \lambda_\alpha  \, e_\alpha$ then
that $X = \big\{   \alpha\in\mathcal{M} :  \phi_\alpha(s)>0  \big\}$.
Thus $ \Delta_\mathcal{C}$ is the disjoint union of the interiors $\mathrm{int}\Delta_X$ for $X\in\mathcal{C}$. 
\end{lem}
We also define the 
 faces of $\Delta_X$ to be $\Delta_Y$ for all $Y\subset X$ where $Y$ has one less element than $X$.

A path or function of time on $\Delta_\mathcal{C}$  can be approximated arbitrarily closely by a piecewise linear path $p(t)\in\Delta_\mathcal{C}$. This is basically a join the dots operation using straight line segments. This makes sense in a computational system, as we can only check the position for a discrete set of time values.

\begin{propos} \label{linseg}
A line segment contained in $\Delta_\mathcal{C}$ must lie entirely within the interior $\mathrm{int}\Delta_X$ for a unique
$X\in\mathcal{C}$, with the exception of the end points of the line segment which may either be in $\mathrm{int}\Delta_X$ or on a face (or intersection of faces) of $\Delta_X$. 
\end{propos}

As a general principle for a state space $S$ and partition of unity $\phi:S\to \Delta_\mathcal{C}$, if we are in state $s\in S$ and $\phi(s)\in \mathrm{int}\Delta_X$ then we should be in mode $X$, as $X$ is the smallest (or simplest) element of 
$\mathcal{C}$ which has $\phi(s)$ in its simplex. As we shall see, things are not quite this simple, however we can use this to see why we only considered transitions to a subset or superset mode. 

\begin{cor} \label{linseg}
Suppose that a line segment contained in $\Delta_\mathcal{C}$ has its beginning in $\mathrm{int}\Delta_X$ for some $X\in \mathcal{C}$, but its end is not in $X\in \mathcal{C}$. Then we have two possibilities 

1) Most of the line segment is contained in $\mathrm{int}\Delta_X$ and its end is in a face (or intersection of faces) of $\Delta_X$. 

2) Most of the line segment is contained in $\mathrm{int}\Delta_Z$ for some $X\subsetneq Z$. 
\end{cor}

Corollary~\ref{linseg} gives us our two cases for transferring from mode $X$. In case (1) we transfer to a subset mode, and in case (2) we transfer to a superset mode. Any more complicated transition would have to be a composition of these two cases. For example we might transfer from $\{\alpha,\beta\}$ to $\{\alpha,\gamma\}$ by first moving to the face 
$\{\alpha\}$ of $\{\alpha,\beta\}$ and then to the superset $\{\alpha,\gamma\}$ of $\{\alpha\}$. 

To summarise, remembering that the distributed algorithms actually have the final say on requesting transfers, we might propose a method depending on testing for zero values.

\begin{figure}
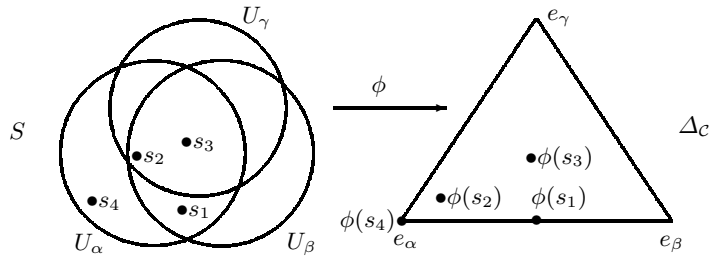

\begin{center}

\unitlength 0.6 mm


\caption{A partition of unity for Example~\ref{youd} }
\label{part789}
\end{center}
\end{figure}

\begin{exam} 
\label{youd}  
The state space $S=U_\alpha\cup U_\beta\cup U_\gamma$ is illustrated in Figure~\ref{part789}, so 
$\mathcal{M}=\{\alpha,\beta,\gamma\}$ and $\mathcal{C}$ consists of all subsets of $\mathcal{M}$ as all the intersections are non-empty. 
We write an element of the real simplical complex  $ \Delta_\mathcal{C}$ in Section~\ref{real_simplicial_complexes} using the basis $\{e_\alpha,e_\beta,e_\gamma\}$ of $\mathbb{R}^\mathcal{M}$ as $a\,e_\alpha+b\,e_\beta+c\,e_\gamma$, or in more compact notation $(a,b,c)\in  \Delta_\mathcal{C}$ for real $a,b,c\ge 0$ with $a+b+c=1$.  Figure~\ref{part789} illustrates four states $s_1,s_2,s_3,s_4$ and their images under a partition of unity $\phi:S\to \Delta_\mathcal{C}$:
\begin{center}
$\phi( s_1)=(\tfrac12,\tfrac12,0)$, 
$\phi(s_2)=(\tfrac{18}{20},\tfrac1{20},\tfrac1{20})$, $\phi(s_3)=(\tfrac2{5},\tfrac2{5},\tfrac1{5})$  and $\phi(s_4)=(1,0,0)$ 
\end{center}

We will switch modes based on a simple test as to whether the coordinates are zero or not. 

For the first case, suppose that the path $s(t)\in S$ as a function of time $t$ begins at $s(0)=s_1$ and proceeds to  $s(1)=s_3$ so that in $\Delta_\mathcal{C}$ the image of the path  $\phi(s(t))$ is a straight line path from  $\phi(s_1)$ to $\phi(s_3)$. Set $\phi(s(t))=(a(t),b(t),c(t))$, so initially $c(0)=0$ and eventually $c(1)=\frac15$. We begin in mode $\{\alpha,\beta\}$ and, based on our instructions for changing modes, at the first value of time $t$ we check that has $c(t)>0$ we transfer to mode  $\{\alpha,\beta,\gamma\}$.

For the second case, suppose that the path $s(t)\in S$ begins at $s(0)=s_3$ and proceeds to  $s(1)=s_1$ so that $\phi(s(t))$ is a straight line path from
$\phi(s_3)$ to $\phi(s_1)$ in $ \Delta_\mathcal{C}$. Now initially $c(0)=\frac15$ and eventually $c(1)=0$. We begin in mode $\{\alpha,\beta,\gamma\}$ and, based on our instructions for changing modes, at the first value of time $t$ we check that has $c(t)=0$ we transfer to mode  $\{\alpha,\beta\}$.
\end{exam}

This example reveals a potential flaw in the method, called the Zeno effect and discussed next in Section~\ref{Zenoav}. 

However, as far as the computable structure discussed in Principles \ref{prin1} and \ref{prin2} is concerned, there are these issues we also need to discuss:

(a) What does the functoriality of the transition functions mean for the data? 

(b) What are the domains of the partial functions implementing the transitions?

(c) The use of the idealised function $\phi:S\to \Delta_\mathcal{C}$ (not in the model) rather than its computable localisations $\phi_X :S\to \Delta_\mathcal{C}$ (in the model).

\subsection{Avoiding Zeno} \label{Zenoav}
 Zeno of Elea's paradoxes involve infinitely many things happening in finite time. In the theory of hybrid systems, Zeno behaviour refers to a system making an infinite number of discrete changes of state (as opposed to a continuous change) in a finite time \cite{Zeno}. This is something to avoid in changing modes, as the control system would simply fail. In general, it is not possible to guarantee this. 
 
 In theory, a perfectly rigid ball bouncing on a perfectly rigid surface will bounce infinitely many times and then come to rest in a finite time. However, as far as any physical measurement is concerned, the ball will effectively come to rest in a finite number of bounces. By imposing a nonzero threshold on the size of bounce that we measure we get around the Zeno behaviour. 
 
 Looking at Example~\ref{youd} as illustrated in Figure~\ref{part789}, we see that if a path `bounces' up and down on the horizontal side of the triangle near the point $\phi(s_1)$ then it will change mode very rapidly, from $\{\alpha,\beta,\gamma\}$ to $\{\alpha,\beta\}$ and back again repeatedly. This is because the threshold level for transition one way is the same as that for the other direction. To get round this effect we need to impose a nonzero difference between the threshold levels.
 
 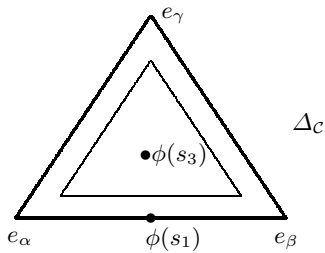
\begin{figure}
\begin{center}

\unitlength 0.6 mm
\begin{picture}(150,50)(44,24)

\linethickness{0.3mm}
\multiput(90,35)(0.12,0.18){250}{\line(0,1){0.18}}
\linethickness{0.3mm}
\multiput(120,80)(0.12,-0.18){250}{\line(0,-1){0.18}}
\linethickness{0.3mm}
\put(90,35){\line(1,0){60}}
\put(155,56){\makebox(0,0)[cc]{$\Delta_\mathcal{C}$}}

\put(91,30.5){\makebox(0,0)[cc]{$e_\alpha$}}
\put(150,30){\makebox(0,0)[cc]{$e_\beta$}}
\put(125,80){\makebox(0,0)[cc]{$e_\gamma$}}
\put(120,35){\makebox(0,0)[cc]{$\bullet$}}


\put(125,30.5){\makebox(0,0)[cc]{$\phi(s_1)$}}


\put(125,49){\makebox(0,0)[cc]{$\bullet \phi(s_3)$}}


\linethickness{0.1mm}
\multiput(100,40)(0.12,0.18){166}{\line(0,1){0.18}}
\linethickness{0.1mm}
\multiput(120,70)(0.12,-0.18){166}{\line(0,-1){0.18}}
\linethickness{0.1mm}
\put(100,40){\line(1,0){40}}

\end{picture}

\caption{Avoiding the Zeno effect in Example~\ref{youd3} }
\label{part78900}
\end{center}
\end{figure}

\begin{exam} \label{youd3}  Continuing Example~\ref{youd} as illustrated in Figure~\ref{part789}, we use different instructions for changing modes. If we begin in the interior of the triangle (2-simplex) $\alpha\beta\gamma$ then we will change mode to a side or a vertex only when we actually touch the boundary of the triangle. However, if we begin from a side or a vertex (a submode of $\{\alpha,\beta,\gamma\}$) we will not change mode to $\{\alpha,\beta,\gamma\}$ unless we cross a boundary strictly inside the interior of the triangle. We visualise this boundary as the inner triangle in Figure~\ref{part78900}. Thus, in the neighbourhood of the points $\phi(s_1)$ and $\phi(s_3)$ in Figure~\ref{part78900}, a transition from  $\{\alpha,\beta,\gamma\}$ to  $\{\alpha,\beta\}$ will happen on hitting the line given by points $(a,b,0)\in \Delta_{\mathcal{C}}$, but the transition from $\{\alpha,\beta\}$ to $\{\alpha,\beta,\gamma\}$ will only happen on hitting the inner horizontal line given by points $(a,b,\eta)\in \Delta_{\mathcal{C}}$ for some fixed $\eta>0$. Now a ball bouncing on the outer horizontal line near $\phi(s_1)$ will only have finitely many bounces higher than the inner horizontal line, so having a threshold value $\eta>0$ aviods the Zeno effect. (No continuous path can cross from one horizontal line to the other infinitely many times in a finite time.)
\end{exam}

\subsection{The importance of functoriality} \label{functorimp}

\begin{figure}
\begin{center}

\unitlength 0.6 mm
\begin{picture}(100,50)(0,15)
\linethickness{0.1mm}
\put(30,20){\line(1,0){70}}
\linethickness{0.1mm}
\multiput(30,20)(0.12,0.17){220}{\line(0,1){0.17}}
\linethickness{0.3mm}
\multiput(36.88,30)(0.32,0.12){167}{\line(1,0){0.32}}
\linethickness{0.3mm}
\multiput(45,20)(0.18,0.12){250}{\line(1,0){0.18}}
\linethickness{0.3mm}
\put(30,20){\line(1,0){15}}
\linethickness{0.3mm}
\multiput(30,20)(0.24,0.12){250}{\line(1,0){0.24}}
\put(97,50){\makebox(0,0)[cc]{$\bullet \ \phi(s_2)$}}

\put(23,20){\makebox(0,0)[cc]{$\phi(s_4)\  \bullet$}}

\linethickness{0.3mm}
\multiput(30,20)(0.12,0.18){57}{\line(0,1){0.18}}
\put(49.38,38.12){\makebox(0,0)[cc]{$c$}}

\put(59.38,25.62){\makebox(0,0)[cc]{$b$}}

\put(38.75,26.88){\makebox(0,0)[cc]{$a$}}

\put(30,16){\makebox(0,0)[cc]{$e_\alpha$}}
\put(100,30){\makebox(0,0)[cc]{$\Delta_\mathcal{C}$}}

\put(34,29){\makebox(0,0)[cc]{$p\  \bullet$}}

\put(45,20){\makebox(0,0)[cc]{$\bullet$}}
\put(45,15){\makebox(0,0)[cc]{$q$}}

\end{picture}

\caption{Small perturbations to paths from Example~\ref{youd9}}
\label{threes}
\end{center}
\end{figure}
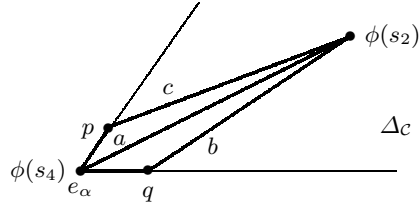

\begin{exam} \label{youd9}
We magnify a portion of Figure~\ref{part789} (which is explained in Example~\ref{youd}) and display the result in 
Figure~\ref{threes}. Then we draw three piecewise linear paths $a,b,c$ from $\phi(s_2)$ to the vertex $\phi(s_4)=e_\alpha$. Suppose that we transfer from mode $\{\alpha,\beta,\gamma\}$ to a subset mode when we touch a side or vertex
(imposing other threshold values would require a shift in the paths but not fundamentally alter the consequences). 

First start in mode $\{\alpha,\beta,\gamma\}$ at $\phi(s_2)$ and move along path $a$. The first point on the boundary we hit is $\phi(s_4)=e_\alpha$, so we invoke the transition function 
${}_{ \{\alpha\}}\mathrm{proj}{}_{\{\alpha,\beta,\gamma\}}:S_{\{\alpha,\beta,\gamma\}}\to S_{ \{\alpha\}}$.

If instead we move along path $b$ then we first move to mode $\{\alpha,\beta\}$ when we get to $q$ and subsequently to mode
$\{\alpha\}$, so we get a composition of two transition functions. 

If instead we move along path $c$ then we first move to mode $\{\alpha,\gamma\}$ when we get to $p$ and subsequently to mode
$\{\alpha\}$, so we get a composition of two transition functions. 

Now we can imagine that $p$ and $q$ are so close to the vertex $e_\alpha$ that there is effectively no time for measurement or calculation between them and reaching the vertex, and the three paths $a,b,c$ would be effectively indistinguishable. To avoid the complication of a discontinuity in the data given by a continuous deformation of the paths we would require that for $p$ and $q$ sufficiently close to the vertex that the transition functions gave the same result, i.e.\
\[
{}_{ \{\alpha\}}\mathrm{proj}{}_{\{\alpha,\beta,\gamma\}} = 
{}_{ \{\alpha\}}\mathrm{proj}{}_{\{\alpha,\gamma\}} \circ {}_{ \{\alpha,\gamma\}}\mathrm{proj}{}_{\{\alpha,\beta,\gamma\}} =
{}_{ \{\alpha\}}\mathrm{proj}{}_{\{\alpha,\beta\}} \circ {}_{ \{\alpha,\beta\}}\mathrm{proj}{}_{\{\alpha,\beta,\gamma\}} \ .
\]

\end{exam}
 
A more general argument than Example~\ref{youd9} would give the general functorial property for $\mathrm{proj}$ 
in Principle~\ref{prin2}, and reversing the paths would give the functorial property for $\mathrm{inc}$ in Principle~\ref{prin1}. (The functorial property is not strictly necessary, but without it the resulting discontinuity when the path varies would complicate a geometric analysis of the system.)

\subsection{The philosophy of the transition functions} \label{funjuphil}
It is important to see how transition functions link to information management. 

%
%

Consider moving from a mode $X$ to a simpler subset mode $Y\subset X$ when we no longer believe that its complement $X\setminus Y$ in $X$ is relevant. In agreement with William of Ockham we establish a principle to move to a simpler (subset) mode where possible\footnote{Frustra fit per plura quod potest fieri per pauciora = It is futile to do with more things that which can be done with fewer. 
William of Ockham \textit{Summa Totius Logicae}, circa 1323 A.D.} . 


%
%

Conversely, consider moving from mode $X$ to a more complicated superset mode. This is 
 $X\subset Z$ is more difficult as we are moving into an unknown country where there are factors in mode $Z$ that mode $X$ was never designed to understand. 
Firstly, we would not want to move from mode $X$ if we believe that it is doing well; there has to be a motivation to move -- a crisis brewing. 
However, not so obviously, there is a problem when this crisis is too large.  It is not that we cannot change mode in a time of great crisis (indeed there may be little alternative), it is rather that we would be so far into an unknown country that the result of doing so would not be predictable. 

%
%

\subsection{The domain of the transition functions} \label{funjup} 

To apply the principles of Section~\ref{funjuphil}, we need to quantify the extent to which a mode or subset of a mode is relevant. We need  functions to measure our `belief' in a mode's fitness for purpose: 
\begin{center}
 $\mathbb{B}(Y) : S \to [0, 1]$ and $\mathbb{B}_X(Y) : S_X \to [0, 1]$ .
\end{center}
The function $\mathbb{B}(Y)$ formalises our \textit{belief}, as an element of $[0,1]$, that the subset $Y\subset\mathcal{M}$ is relevant in state $s\in S$. In addition to the abstract picture for  $s\in S$ we have $\mathbb{B}_X(Y)$ as our computable approximation in mode $X$, using $\phi_X:S_X\to 
\Delta_{ \mathcal{C}  }$ and $\tilde s\in S_X$. 
Rather than use a general theory of belief and evidence (e.g.,\ \cite{TheoryOfEvidence}), referring to the notation in Definition~\ref{realpu} we can simply use the sum
\begin{align} \label{rousp}
\mathbb{B}(Y)(s) = \sum_{ \alpha\in Y} \phi_\alpha(s) \ ,\quad \mathbb{B}_X(Y)(\tilde s) = \sum_{ \alpha\in Y} \phi_{X\alpha}(\tilde s)\ .
\end{align}

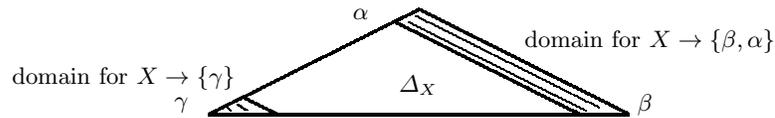
\begin{figure}
\begin{center}

\unitlength 0.7 mm
\begin{picture}(100.62,30)(0,15)
\linethickness{0.3mm}
\put(20,20){\line(1,0){80}}
\linethickness{0.3mm}
\multiput(20,20)(0.24,0.12){167}{\line(1,0){0.24}}
\linethickness{0.3mm}
\multiput(60,40)(0.24,-0.12){167}{\line(1,0){0.24}}
\linethickness{0.3mm}
\multiput(55.62,37.5)(0.24,-0.12){146}{\line(1,0){0.24}}
\linethickness{0.3mm}
\multiput(26.88,23.12)(0.24,-0.12){26}{\line(1,0){0.24}}
\linethickness{0.1mm}
\multiput(25.62,21.88)(0.19,-0.12){10}{\line(1,0){0.19}}
\linethickness{0.1mm}
\multiput(23.75,21.25)(0.12,-0.12){5}{\line(1,0){0.12}}
\linethickness{0.1mm}
\multiput(59.38,38.75)(0.24,-0.12){146}{\line(1,0){0.24}}
\linethickness{0.1mm}
\multiput(58.12,37.5)(0.24,-0.12){141}{\line(1,0){0.24}}
\put(49,39){\makebox(0,0)[cc]{$\alpha$}}

\put(103,21.88){\makebox(0,0)[cc]{$\beta$}}

\put(15,21.88){\makebox(0,0)[cc]{$\gamma$}}

\put(4,26.88){\makebox(0,0)[cc]{domain for $X\to\{\gamma\}$}}

\put(104,34){\makebox(0,0)[cc]{domain for $X\to\{\beta,\alpha\}$}}

\put(60,25.62){\makebox(0,0)[cc]{$\Delta_X$}}

\end{picture}

\caption{Illustration of the domains of the transition functions ${}_Y\mathrm{proj}{}_X:S_X\to S_Y$
for $X=\{\alpha,\beta,\gamma\}$}
\label{thou}
\end{center}
\end{figure}

Consider the two forms of transition.
\newline
\newline
\noindent \textbf{\textit{Moving from a mode $X$ to a subset mode $Y\subset X$}}. 
In Examples~\ref{youd3}  and \ref{youd9} we transferred from mode $X$ to $Y\subset X$ when our path in $\Delta_X$ touched the boundary. When we are at a point in the boundary $\psi_X(\tilde s)\in \Delta_Y\subset \Delta_X$ then by definition $\mathbb{B}_X(X\setminus Y)(\tilde s) =0$. However, as the initiation of mode transfer is the responsibility of the algorithm $\mathcal{B}_X$ it is best to allow some \textit{wiggle room} by a small parameter $\epsilon_{X\to Y}$.
The domain is illustrated in Figure~\ref{thou}.

\begin{defin} \label{projDom} For modes $Y\subset X$ the domain of the transition function ${}_Y\mathrm{proj}{}_X:S_X\to S_Y$ includes the following set, for a given $\epsilon_{X\to Y}>0$:
\[
\big\{  \tilde s\in S_X :    \mathbb{B}_X(X\setminus Y)(\tilde s)     <\epsilon_{X\to Y}\big\}.
\]
\end{defin}

\begin{figure}
\begin{center}

\unitlength 0.7 mm
\begin{picture}(120,20)(0,25)
\linethickness{0.4mm}
\put(10,30){\line(1,0){110}}
\linethickness{0.3mm}
\put(10,35){\line(1,0){50}}
\put(60,35){\vector(1,0){0.12}}
\put(10,35){\vector(-1,0){0.12}}
\linethickness{0.3mm}
\put(60,37.5){\line(1,0){30}}
\put(90,37.5){\vector(1,0){0.12}}
\put(60,37.5){\vector(-1,0){0.12}}
\linethickness{0.3mm}
\put(90,35){\line(1,0){30}}
\put(120,35){\vector(1,0){0.12}}
\put(90,35){\vector(-1,0){0.12}}
\put(90,30){\makebox(0,0)[cc]{$\bullet$}}

\put(60,30){\makebox(0,0)[cc]{$\bullet$}}

\put(105,40){\makebox(0,0)[cc]{$\kappa$omfort zone}}

\put(32.5,42.5){\makebox(0,0)[cc]{}}

\put(75,42){\makebox(0,0)[cc]{growing crisis}}

\put(120,30){\makebox(0,0)[cc]{$\bullet$}}

\put(10,30){\makebox(0,0)[cc]{$\bullet$}}

\put(35,25){\makebox(0,0)[cc]{$\mathbb{B}_X(X)(\tilde s)$}}

\put(10,25){\makebox(0,0)[cc]{$0$}}

\put(60,25){\makebox(0,0)[cc]{$\pi_X$}}

\put(90,25){\makebox(0,0)[cc]{$\kappa_X$}}

\put(120,25){\makebox(0,0)[cc]{$1$}}

\put(32.5,40){\makebox(0,0)[cc]{$\pi$anic zone}}

\end{picture}

\caption{The comfort and panic zones for a mode $X$ in state $\tilde s\in S_X$}
\label{thuu}
\end{center}
\end{figure}
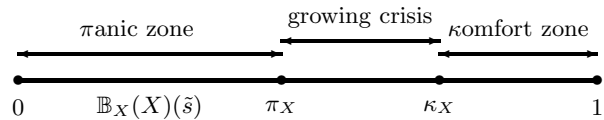

\noindent \textbf{\textit{Moving from a mode $X$ to a superset mode $X\subset Z$}}. 
In mode $X$ in state $\tilde s\in S_X$ the number $\mathbb{B}_X(X)(\tilde s)$ measures how confident mode $X$ is that it models state $\tilde  s\in S_X$ satisfactorily. We highlight two subjective values of belief, the values
$0<\pi_X<\kappa_X<1$ where $\kappa_X$ bounds the {\em $\kappa$omfort zone} and $\pi_X$ the {\em $\pi$anic zone}
as in Figure~\ref{thuu}. 

If we are in the comfort zone for mode $X$ then we are happy to stay in mode $X$ (cf.  Ockham's razor: we do not move to a superset unless we have to). 

In the `growing crisis' zone we look for a superset mode to move to. 

In the panic zone the search  becomes urgent as there is serious doubt about how effective mode $X$ is at controlling the system. 

However, for any particular $X\subset Z$ the domain of the transition function ${}_Z\mathrm{inc}{}_X:S_X\to S_Z$ is rather more difficult, as we are moving into an unknown country where there are factors in mode $Z$ that mode $X$ was never designed to understand. 

One obvious condition is that we would like a definite reason to have all the new elements in $Z$ (Ockham's razor again), so we could ask for 
$\phi_{X\beta}(\tilde s)>0$ for all $\beta\in Z\setminus X$. 

Another obvious condition is that we would prefer to jump into the comfort zone of mode $Z$, as there might be little point in changing from one crisis to another.

\begin{defin} \label{incDom} For $X\subset Z$ the domain of the transition function  ${}_Z\mathrm{inc}{}_X:S_X\to S_Z$  includes
\[
\big\{  \tilde s\in S_X : \pi_{X}  \le \mathbb{B}_X(X)(\tilde s)  \ \& \ 
\kappa_{X}  \le \mathbb{B}_X(Z)(\tilde s)
 \ \& \ 
 \phi_{X\beta}(\tilde s)>0 \ \mathrm{for\ all} \  \beta\in Z\setminus X
  \big\}
\] 
\end{defin}

\noindent To summarise this, we can move out of $X$ if 

(i) we are not in the panic zone for $X$; 

(ii) if we can move into the comfort zone for the new mode $Z$ (or at least what $X$ estimates is the comfort zone for $Z$); and 

(iii) if each new element $\beta\in Z$ is justified by $ \phi_{X\beta}(\tilde s)>0$. 

\noindent Note that we have allowed the possibility of moving from a comfort zone to a comfort zone if the algorithm wishes. 

%

\subsection{The existence of $S$ and the consistency of the transition functions} \label{conjup}

To shape our narrative, we have spoken about the idealisations of the global state space $S$ and the partition of unity $\phi:S\to\Delta_\mathcal{C}$. However, our theoretical model of a system is based on the idea that to compute we have \textit{only} local sets of states $S_X$ and computable functions $\phi_X:S_X\to\Delta_\mathcal{C}$ that belong to the modes. Actually, this is the \textit{raison d'$\hat{e}$tre} of modes: modes address the problem that $S$ may not exist in any meaningful sense.  Whilst a simple physical system may have a global $S$ as a workable mathematical abstraction, what is $S$ when we have an autonomous vehicle in a city?\footnote{Just as complex, what could be $S$ in some multiple agency social services situation? cf. \ref{scope}}

All we have are the local structures $S_X$, and any meaning for the global system depends upon gluing the local modes together. However, there are consistency issues which arise for the gluing procedure, which in our case leads to criteria on the transition functions. 

Our first stage in gluing the modes together is understanding that the algorithm may no longer be right for the behaviour of the system: if the algorithm is going wrong, then at least it realises that it may be going wrong. 

\begin{exam} \label{Pendulum}
A control system in mode $X$ involving a pendulum is written using the formula for a small amplitude oscillation and simple harmonic motion. During the operation of the system, energy is fed into the pendulum and its amplitude of oscillation increases significantly. As this happens the formula will fail to give a good approximation to the motion and the whole control system may collapse. What we need is that the algorithm can flag up that it is running into problems, and our standard method of setting such a flag is through the value of $\phi_X$ on the state
$\tilde s\in S_X$. 
\end{exam}

\begin{principle} \label{prin9}
Following Figure~\ref{thuu} we  assume that 
\[
\mathcal{W}_X=
\big\{  \tilde s\in S_X :  \pi_{X}  \le \mathbb{B}_X(X)(\tilde s)  \big\} 
\]
is a set of states which is modelled `reasonably well' by mode $X$, where $\pi_X\in (0,1)$ is the $\pi$anic level. By modelled `reasonably well' we assume that for a state $\tilde s\in S_X$ a computation carried out on the system by the algorithm for mode $X$ is likely to give a good answer, but no such guarantee exists outside $\mathcal{W}_X$.
\end{principle}

One may ask: Why we do not simply restrict the local states to this set in the first place, as then all our calculations would be `good'.  The reason is for many realworld stories where something has gone wrong, the path to safety has been through possibilities rather than certainties. By deleting all but guaranteed options, we merely make it more likely that we will run out of options. 
Stated alternatively in terms of our hypothetical state space $S$, if we reduce the size of the sets in a cover we may not get a cover.

The next thing to do is to examine the consistency of the functions $\phi_X$ for $\mathrm{inc}$ (see Definition~\ref{incDom}) and $\mathrm{proj}$ (see Definition~\ref{projDom}). Of course, this compatibility will only be expected to work in a subset of $\mathcal{W}_X$ (see Principle~\ref{prin9}) as all our calculations are suspect outside that set. Compatibility corresponds to the following diagram, commuting on the specified domains intersected with $\mathcal{W}_X$
\begin{align} \label{compatDiag}
 \UseComputerModernTips
  \xymatrix{
S_X  \ar[r]^{\phi_X}  \ar[d]_{ {}_Z\mathrm{inc}{}_X  / {}_Z\mathrm{proj}{}_X } &  \Delta_\mathcal{C}  
  \\
S_Z \ar[ur]_{\phi_Z}   &  
}
\end{align}
Finally we come the composition ${}_X\mathrm{proj}_Z\circ {}_Z\mathrm{inc}_X:S_X\to S_X$ for $X\subset Z$ in Principle~\ref{prin3}. 
Definition~\ref{incDom} gives the guaranteed domain of ${}_Z\mathrm{inc}{}_X:S_X\to S_Z$, and this is a subset of $\mathcal{W}_X$, namely
\[
\big\{  \tilde s\in S_X : \pi_{X}  \le \mathbb{B}_X(X)(\tilde s)  \ \mathrm{and}\ 
\kappa_{X}  \le \mathbb{B}_X(Z)(\tilde s)
 \ \mathrm{and}\ 
 \phi_X(\tilde s)_\beta>0 \ \mathrm{for\ all} \  \beta\in Z\setminus X
  \big\}
\]
so we deduce that 
\[
 \mathbb{B}_X(Z\setminus X)(\tilde s)= \mathbb{B}_X(Z)(\tilde s)- \mathbb{B}_X( X)(\tilde s) \le 1 - \pi_X
\]
so by the consistency of $\phi_X$ and $\phi_Z$ in (\ref{compatDiag}) we have for
$\tilde r= {}_Z\mathrm{inc}_X(\tilde s)\in S_Z$ that $ \mathbb{B}_Z(Z\setminus X)(\tilde r)\le 1 - \pi_X$.
By Definition~\ref{projDom}, the image is contained in 
 the domain of ${}_X\mathrm{proj}{}_Z:S_Z\to S_X$ if $1-\pi_X<\epsilon_{Z\to X}$. 

Now, in this case the condition that the composition is the identity would make formal sense, and even if the domains and images did not quite match there might be a restriction of the domain on which it made sense. However, similarly to the functoriality discussed in Section~\ref{functorimp}, we should regard this condition as one which makes reasoning about the system easier, and therefore should be a condition which is true as far as possible, rather than a condition where we have to mess with everything just to make sure that it applies exactly.

%

It may be worth pausing and asking just what is the model of modes mathmetically. In topology, a functor from a category, whose objects are given by a cover of a topological space, to another category is called a presheaf (see Definition~\ref{defmor}). We assume a fixed concrete cover of $S$ and the functorial conditions in Section~\ref{functorimp}, which may not always be exact, to make a similar idea. 
Using the compatibility conditions (\ref{compatDiag}) and the one sided identity condition of Principle~\ref{prin3} we try to glue the objects for each element of the cover together as seamlessly as possible, and using a further slight abuse of notation we might call this a sheaf construction. The model of modes is mathematically this:

\begin{defin} \label{sheaf} 
A {\em{abstract simplicial sheaf datatype}} is a simplical complex $\mathcal{C}$ with a local computable structure as given in Section~\ref{compstr} obeying (to a reasonable extent) the functorial conditions in Section~\ref{functorimp}, the one sided identity condition of Principle~\ref{prin3} and the compatibility conditions (\ref{compatDiag}).
\end{defin}

\subsection{A summary in preparation for a case study} \label{summ}
In describing the model we have made a distinction in our theorising between the real physical system and its operating environment and the digital system that is designed to approximate and control it.
\\
\\
\textbf{Real world idealisation.} Consider a complex system that must perform various tasks. To have an effective mathematical descriptions and appropriate algorithms for these tasks, the system is designed in pieces, which must be combined to specify the system. These pieces we have called \textit{modes}. The patchwork of modes is formalised by imagining global state space description $S$ for the whole system and postulating idea of a cover of the state space $S$. This cover determines all the local state spaces that are the basis of the modes. The key techincal idea is that these local state spaces, and the modes that depend upon them, have the structure of an \textit{abstract simplical complex} $\mathcal{C}$. The modes are represented by simplices. The actual control of the system is achieved by \textit{sensors} and \textit{actuators}. The flow within the system is given by conditions for \textit{transitions} between modes. The most important ingredient of these conditions is the \textit{partition of unity} $\phi:S\to\Delta_\mathcal{C}$ which
determines the fitness of a mode $X$ to control the system in state $s\in S$. 
\\
\\
\textbf{Computable world  idealisation.} The programmers construct sets $S_X$ which hold the data for the picture of reality available to the mode $X$. The control of the system in mode $X$ is implemented by an algorithm $\mathcal{B}_X$ written using a many sorted algebra $\mathcal{A}_X$. The communication with the real world sensors and actuators is performed by oracles, which are formally functions in $\mathcal{A}_X$. The local representation of the partition of unity $\phi_X:S_X\to \Delta_\mathcal{C}$ is a function in $\mathcal{A}_X$.  Two special components effect the performance of modes.

\textit{The Confidence Levels.} To judge the effectiveness of a mode $X$ to control the system, using $\phi_X$
 as shown in Figure~\ref{thuu} we split 
into a \textit{comfort zone}  (i.e.\ above  $\kappa_{X}$ mode $X$ is doing a `good' job)
and an \textit{panic zone} 
 (i.e.,\ below the level $\pi_{X}$ mode $X$ may not control the system sufficiently well). 

\textit{Transitions between modes.} Transitions are given by partial functions ${}_Y\mathrm{proj}{}_X:S_X\to S_Y$ for $Y\subset X$ and ${}_Z\mathrm{inc}{}_X:S_X\to S_Z$ for $X\subset Z$, and transition is initiated by the call of a formal function in $\mathcal{A}_X$ by the algorithm. Their (minimal) domains are described in Definition~\ref{incDom} and  Definition~\ref{projDom}. There is a consistency condition between these transition functions and $\phi_X$ described in 
(\ref{compatDiag}). 

A desirable condition, but not essential, on the transition functions is that they are functorial on a reasonable domain (see Principle~\ref{prin1} and Principle~\ref{prin2}), and that we have an one sided inverse property as in 
Principle~\ref{prin3}. 


\section{Case Study: Autonomous Racing Cars} \label{autonomous_racing_cars}

We give an example to illustrate the components of the model. To do this within the confines of a paper we choose a simple example for ease of understanding and to suggest how the example scales up.  As a result, it is easy to grasp the whole system and the transition functions from one mode to another are chosen to be rather trivial. We start with a simple track and car and then introduce a second car and a chicane.

\subsection{Single cars and the presheaf}
\begin{figure}
\begin{center}
\unitlength 0.7 mm
\begin{picture}(115,40)(10,29)
\linethickness{0.3mm}
\put(20,60){\line(0,1){10}}
\linethickness{0.3mm}
\put(20,60){\line(1,0){70}}
\linethickness{0.3mm}
\put(20,70){\line(1,0){70}}
\linethickness{0.3mm}
\linethickness{0.3mm}
\linethickness{0.3mm}
\linethickness{0.3mm}
\linethickness{0.3mm}
\linethickness{0.3mm}
\linethickness{0.3mm}
\linethickness{0.3mm}

\linethickness{0.3mm}
\multiput(90,40)(0.5,0.01){1}{\line(1,0){0.5}}
\multiput(90.5,40.01)(0.5,0.04){1}{\line(1,0){0.5}}
\multiput(91,40.05)(0.49,0.06){1}{\line(1,0){0.49}}
\multiput(91.49,40.11)(0.49,0.09){1}{\line(1,0){0.49}}
\multiput(91.98,40.2)(0.49,0.11){1}{\line(1,0){0.49}}
\multiput(92.47,40.31)(0.48,0.14){1}{\line(1,0){0.48}}
\multiput(92.95,40.44)(0.47,0.16){1}{\line(1,0){0.47}}
\multiput(93.42,40.6)(0.23,0.09){2}{\line(1,0){0.23}}
\multiput(93.88,40.79)(0.23,0.1){2}{\line(1,0){0.23}}
\multiput(94.34,40.99)(0.22,0.11){2}{\line(1,0){0.22}}
\multiput(94.78,41.22)(0.22,0.12){2}{\line(1,0){0.22}}
\multiput(95.21,41.47)(0.21,0.14){2}{\line(1,0){0.21}}
\multiput(95.63,41.74)(0.2,0.15){2}{\line(1,0){0.2}}
\multiput(96.04,42.03)(0.13,0.1){3}{\line(1,0){0.13}}
\multiput(96.43,42.34)(0.12,0.11){3}{\line(1,0){0.12}}
\multiput(96.8,42.67)(0.12,0.12){3}{\line(1,0){0.12}}
\multiput(97.16,43.02)(0.11,0.12){3}{\line(0,1){0.12}}
\multiput(97.5,43.38)(0.11,0.13){3}{\line(0,1){0.13}}
\multiput(97.82,43.77)(0.1,0.13){3}{\line(0,1){0.13}}
\multiput(98.12,44.16)(0.14,0.21){2}{\line(0,1){0.21}}
\multiput(98.4,44.57)(0.13,0.21){2}{\line(0,1){0.21}}
\multiput(98.66,45)(0.12,0.22){2}{\line(0,1){0.22}}
\multiput(98.9,45.44)(0.11,0.22){2}{\line(0,1){0.22}}
\multiput(99.12,45.89)(0.1,0.23){2}{\line(0,1){0.23}}
\multiput(99.31,46.35)(0.17,0.47){1}{\line(0,1){0.47}}
\multiput(99.48,46.82)(0.15,0.48){1}{\line(0,1){0.48}}
\multiput(99.63,47.29)(0.12,0.48){1}{\line(0,1){0.48}}
\multiput(99.75,47.77)(0.1,0.49){1}{\line(0,1){0.49}}
\multiput(99.85,48.26)(0.07,0.49){1}{\line(0,1){0.49}}
\multiput(99.92,48.76)(0.05,0.5){1}{\line(0,1){0.5}}
\multiput(99.97,49.25)(0.02,0.5){1}{\line(0,1){0.5}}
\put(100,49.75){\line(0,1){0.5}}
\multiput(99.97,50.75)(0.02,-0.5){1}{\line(0,-1){0.5}}
\multiput(99.92,51.24)(0.05,-0.5){1}{\line(0,-1){0.5}}
\multiput(99.85,51.74)(0.07,-0.49){1}{\line(0,-1){0.49}}
\multiput(99.75,52.23)(0.1,-0.49){1}{\line(0,-1){0.49}}
\multiput(99.63,52.71)(0.12,-0.48){1}{\line(0,-1){0.48}}
\multiput(99.48,53.18)(0.15,-0.48){1}{\line(0,-1){0.48}}
\multiput(99.31,53.65)(0.17,-0.47){1}{\line(0,-1){0.47}}
\multiput(99.12,54.11)(0.1,-0.23){2}{\line(0,-1){0.23}}
\multiput(98.9,54.56)(0.11,-0.22){2}{\line(0,-1){0.22}}
\multiput(98.66,55)(0.12,-0.22){2}{\line(0,-1){0.22}}
\multiput(98.4,55.43)(0.13,-0.21){2}{\line(0,-1){0.21}}
\multiput(98.12,55.84)(0.14,-0.21){2}{\line(0,-1){0.21}}
\multiput(97.82,56.23)(0.1,-0.13){3}{\line(0,-1){0.13}}
\multiput(97.5,56.62)(0.11,-0.13){3}{\line(0,-1){0.13}}
\multiput(97.16,56.98)(0.11,-0.12){3}{\line(0,-1){0.12}}
\multiput(96.8,57.33)(0.12,-0.12){3}{\line(1,0){0.12}}
\multiput(96.43,57.66)(0.12,-0.11){3}{\line(1,0){0.12}}
\multiput(96.04,57.97)(0.13,-0.1){3}{\line(1,0){0.13}}
\multiput(95.63,58.26)(0.2,-0.15){2}{\line(1,0){0.2}}
\multiput(95.21,58.53)(0.21,-0.14){2}{\line(1,0){0.21}}
\multiput(94.78,58.78)(0.22,-0.12){2}{\line(1,0){0.22}}
\multiput(94.34,59.01)(0.22,-0.11){2}{\line(1,0){0.22}}
\multiput(93.88,59.21)(0.23,-0.1){2}{\line(1,0){0.23}}
\multiput(93.42,59.4)(0.23,-0.09){2}{\line(1,0){0.23}}
\multiput(92.95,59.56)(0.47,-0.16){1}{\line(1,0){0.47}}
\multiput(92.47,59.69)(0.48,-0.14){1}{\line(1,0){0.48}}
\multiput(91.98,59.8)(0.49,-0.11){1}{\line(1,0){0.49}}
\multiput(91.49,59.89)(0.49,-0.09){1}{\line(1,0){0.49}}
\multiput(91,59.95)(0.49,-0.06){1}{\line(1,0){0.49}}
\multiput(90.5,59.99)(0.5,-0.04){1}{\line(1,0){0.5}}
\multiput(90,60)(0.5,-0.01){1}{\line(1,0){0.5}}

\linethickness{0.3mm}
\multiput(90,30)(0.5,0.01){1}{\line(1,0){0.5}}
\multiput(90.5,30.01)(0.5,0.02){1}{\line(1,0){0.5}}
\multiput(91,30.02)(0.5,0.03){1}{\line(1,0){0.5}}
\multiput(91.49,30.06)(0.5,0.04){1}{\line(1,0){0.5}}
\multiput(91.99,30.1)(0.5,0.06){1}{\line(1,0){0.5}}
\multiput(92.49,30.16)(0.49,0.07){1}{\line(1,0){0.49}}
\multiput(92.98,30.22)(0.49,0.08){1}{\line(1,0){0.49}}
\multiput(93.47,30.3)(0.49,0.09){1}{\line(1,0){0.49}}
\multiput(93.96,30.4)(0.49,0.1){1}{\line(1,0){0.49}}
\multiput(94.45,30.5)(0.48,0.12){1}{\line(1,0){0.48}}
\multiput(94.94,30.62)(0.48,0.13){1}{\line(1,0){0.48}}
\multiput(95.42,30.75)(0.48,0.14){1}{\line(1,0){0.48}}
\multiput(95.9,30.89)(0.47,0.15){1}{\line(1,0){0.47}}
\multiput(96.37,31.04)(0.47,0.16){1}{\line(1,0){0.47}}
\multiput(96.84,31.21)(0.47,0.18){1}{\line(1,0){0.47}}
\multiput(97.31,31.38)(0.23,0.09){2}{\line(1,0){0.23}}
\multiput(97.77,31.57)(0.23,0.1){2}{\line(1,0){0.23}}
\multiput(98.23,31.77)(0.23,0.11){2}{\line(1,0){0.23}}
\multiput(98.68,31.98)(0.22,0.11){2}{\line(1,0){0.22}}
\multiput(99.12,32.2)(0.22,0.12){2}{\line(1,0){0.22}}
\multiput(99.57,32.44)(0.22,0.12){2}{\line(1,0){0.22}}
\multiput(100,32.68)(0.21,0.13){2}{\line(1,0){0.21}}
\multiput(100.43,32.93)(0.21,0.13){2}{\line(1,0){0.21}}
\multiput(100.85,33.2)(0.21,0.14){2}{\line(1,0){0.21}}
\multiput(101.27,33.48)(0.2,0.14){2}{\line(1,0){0.2}}
\multiput(101.67,33.76)(0.2,0.15){2}{\line(1,0){0.2}}
\multiput(102.08,34.06)(0.13,0.1){3}{\line(1,0){0.13}}
\multiput(102.47,34.36)(0.13,0.11){3}{\line(1,0){0.13}}
\multiput(102.86,34.68)(0.13,0.11){3}{\line(1,0){0.13}}
\multiput(103.23,35)(0.12,0.11){3}{\line(1,0){0.12}}
\multiput(103.6,35.34)(0.12,0.11){3}{\line(1,0){0.12}}
\multiput(103.96,35.68)(0.12,0.12){3}{\line(0,1){0.12}}
\multiput(104.32,36.04)(0.11,0.12){3}{\line(0,1){0.12}}
\multiput(104.66,36.4)(0.11,0.12){3}{\line(0,1){0.12}}
\multiput(105,36.77)(0.11,0.13){3}{\line(0,1){0.13}}
\multiput(105.32,37.14)(0.11,0.13){3}{\line(0,1){0.13}}
\multiput(105.64,37.53)(0.1,0.13){3}{\line(0,1){0.13}}
\multiput(105.94,37.92)(0.15,0.2){2}{\line(0,1){0.2}}
\multiput(106.24,38.33)(0.14,0.2){2}{\line(0,1){0.2}}
\multiput(106.52,38.73)(0.14,0.21){2}{\line(0,1){0.21}}
\multiput(106.8,39.15)(0.13,0.21){2}{\line(0,1){0.21}}
\multiput(107.07,39.57)(0.13,0.21){2}{\line(0,1){0.21}}
\multiput(107.32,40)(0.12,0.22){2}{\line(0,1){0.22}}
\multiput(107.56,40.43)(0.12,0.22){2}{\line(0,1){0.22}}
\multiput(107.8,40.88)(0.11,0.22){2}{\line(0,1){0.22}}
\multiput(108.02,41.32)(0.11,0.23){2}{\line(0,1){0.23}}
\multiput(108.23,41.77)(0.1,0.23){2}{\line(0,1){0.23}}
\multiput(108.43,42.23)(0.09,0.23){2}{\line(0,1){0.23}}
\multiput(108.62,42.69)(0.18,0.47){1}{\line(0,1){0.47}}
\multiput(108.79,43.16)(0.16,0.47){1}{\line(0,1){0.47}}
\multiput(108.96,43.63)(0.15,0.47){1}{\line(0,1){0.47}}
\multiput(109.11,44.1)(0.14,0.48){1}{\line(0,1){0.48}}
\multiput(109.25,44.58)(0.13,0.48){1}{\line(0,1){0.48}}
\multiput(109.38,45.06)(0.12,0.48){1}{\line(0,1){0.48}}
\multiput(109.5,45.55)(0.1,0.49){1}{\line(0,1){0.49}}
\multiput(109.6,46.04)(0.09,0.49){1}{\line(0,1){0.49}}
\multiput(109.7,46.53)(0.08,0.49){1}{\line(0,1){0.49}}
\multiput(109.78,47.02)(0.07,0.49){1}{\line(0,1){0.49}}
\multiput(109.84,47.51)(0.06,0.5){1}{\line(0,1){0.5}}
\multiput(109.9,48.01)(0.04,0.5){1}{\line(0,1){0.5}}
\multiput(109.94,48.51)(0.03,0.5){1}{\line(0,1){0.5}}
\multiput(109.98,49)(0.02,0.5){1}{\line(0,1){0.5}}
\multiput(109.99,49.5)(0.01,0.5){1}{\line(0,1){0.5}}
\multiput(109.99,50.5)(0.01,-0.5){1}{\line(0,-1){0.5}}
\multiput(109.98,51)(0.02,-0.5){1}{\line(0,-1){0.5}}
\multiput(109.94,51.49)(0.03,-0.5){1}{\line(0,-1){0.5}}
\multiput(109.9,51.99)(0.04,-0.5){1}{\line(0,-1){0.5}}
\multiput(109.84,52.49)(0.06,-0.5){1}{\line(0,-1){0.5}}
\multiput(109.78,52.98)(0.07,-0.49){1}{\line(0,-1){0.49}}
\multiput(109.7,53.47)(0.08,-0.49){1}{\line(0,-1){0.49}}
\multiput(109.6,53.96)(0.09,-0.49){1}{\line(0,-1){0.49}}
\multiput(109.5,54.45)(0.1,-0.49){1}{\line(0,-1){0.49}}
\multiput(109.38,54.94)(0.12,-0.48){1}{\line(0,-1){0.48}}
\multiput(109.25,55.42)(0.13,-0.48){1}{\line(0,-1){0.48}}
\multiput(109.11,55.9)(0.14,-0.48){1}{\line(0,-1){0.48}}
\multiput(108.96,56.37)(0.15,-0.47){1}{\line(0,-1){0.47}}
\multiput(108.79,56.84)(0.16,-0.47){1}{\line(0,-1){0.47}}
\multiput(108.62,57.31)(0.18,-0.47){1}{\line(0,-1){0.47}}
\multiput(108.43,57.77)(0.09,-0.23){2}{\line(0,-1){0.23}}
\multiput(108.23,58.23)(0.1,-0.23){2}{\line(0,-1){0.23}}
\multiput(108.02,58.68)(0.11,-0.23){2}{\line(0,-1){0.23}}
\multiput(107.8,59.12)(0.11,-0.22){2}{\line(0,-1){0.22}}
\multiput(107.56,59.57)(0.12,-0.22){2}{\line(0,-1){0.22}}
\multiput(107.32,60)(0.12,-0.22){2}{\line(0,-1){0.22}}
\multiput(107.07,60.43)(0.13,-0.21){2}{\line(0,-1){0.21}}
\multiput(106.8,60.85)(0.13,-0.21){2}{\line(0,-1){0.21}}
\multiput(106.52,61.27)(0.14,-0.21){2}{\line(0,-1){0.21}}
\multiput(106.24,61.67)(0.14,-0.2){2}{\line(0,-1){0.2}}
\multiput(105.94,62.08)(0.15,-0.2){2}{\line(0,-1){0.2}}
\multiput(105.64,62.47)(0.1,-0.13){3}{\line(0,-1){0.13}}
\multiput(105.32,62.86)(0.11,-0.13){3}{\line(0,-1){0.13}}
\multiput(105,63.23)(0.11,-0.13){3}{\line(0,-1){0.13}}
\multiput(104.66,63.6)(0.11,-0.12){3}{\line(0,-1){0.12}}
\multiput(104.32,63.96)(0.11,-0.12){3}{\line(0,-1){0.12}}
\multiput(103.96,64.32)(0.12,-0.12){3}{\line(0,-1){0.12}}
\multiput(103.6,64.66)(0.12,-0.11){3}{\line(1,0){0.12}}
\multiput(103.23,65)(0.12,-0.11){3}{\line(1,0){0.12}}
\multiput(102.86,65.32)(0.13,-0.11){3}{\line(1,0){0.13}}
\multiput(102.47,65.64)(0.13,-0.11){3}{\line(1,0){0.13}}
\multiput(102.08,65.94)(0.13,-0.1){3}{\line(1,0){0.13}}
\multiput(101.67,66.24)(0.2,-0.15){2}{\line(1,0){0.2}}
\multiput(101.27,66.52)(0.2,-0.14){2}{\line(1,0){0.2}}
\multiput(100.85,66.8)(0.21,-0.14){2}{\line(1,0){0.21}}
\multiput(100.43,67.07)(0.21,-0.13){2}{\line(1,0){0.21}}
\multiput(100,67.32)(0.21,-0.13){2}{\line(1,0){0.21}}
\multiput(99.57,67.56)(0.22,-0.12){2}{\line(1,0){0.22}}
\multiput(99.12,67.8)(0.22,-0.12){2}{\line(1,0){0.22}}
\multiput(98.68,68.02)(0.22,-0.11){2}{\line(1,0){0.22}}
\multiput(98.23,68.23)(0.23,-0.11){2}{\line(1,0){0.23}}
\multiput(97.77,68.43)(0.23,-0.1){2}{\line(1,0){0.23}}
\multiput(97.31,68.62)(0.23,-0.09){2}{\line(1,0){0.23}}
\multiput(96.84,68.79)(0.47,-0.18){1}{\line(1,0){0.47}}
\multiput(96.37,68.96)(0.47,-0.16){1}{\line(1,0){0.47}}
\multiput(95.9,69.11)(0.47,-0.15){1}{\line(1,0){0.47}}
\multiput(95.42,69.25)(0.48,-0.14){1}{\line(1,0){0.48}}
\multiput(94.94,69.38)(0.48,-0.13){1}{\line(1,0){0.48}}
\multiput(94.45,69.5)(0.48,-0.12){1}{\line(1,0){0.48}}
\multiput(93.96,69.6)(0.49,-0.1){1}{\line(1,0){0.49}}
\multiput(93.47,69.7)(0.49,-0.09){1}{\line(1,0){0.49}}
\multiput(92.98,69.78)(0.49,-0.08){1}{\line(1,0){0.49}}
\multiput(92.49,69.84)(0.49,-0.07){1}{\line(1,0){0.49}}
\multiput(91.99,69.9)(0.5,-0.06){1}{\line(1,0){0.5}}
\multiput(91.49,69.94)(0.5,-0.04){1}{\line(1,0){0.5}}
\multiput(91,69.98)(0.5,-0.03){1}{\line(1,0){0.5}}
\multiput(90.5,69.99)(0.5,-0.02){1}{\line(1,0){0.5}}
\multiput(90,70)(0.5,-0.01){1}{\line(1,0){0.5}}

\linethickness{0.3mm}
\put(90,30){\line(0,1){10}}
\put(79,37){\makebox(0,0)[cc]{Finish F}}
\put(79,32){\makebox(0,0)[cc]{$x=L$}}
\put(10,67){\makebox(0,0)[cc]{Start B}}
\put(10,62){\makebox(0,0)[cc]{$x=0$}}

\put(93,73){\makebox(0,0)[cc]{$x=L/2$}}

\put(122,50){\makebox(0,0)[cc]{Curve Cu}}

\put(58,55){\makebox(0,0)[cc]{Straight Str}}



\end{picture}

\caption{A track for racing cars}
\label{chic}
\end{center}
\vspace{-0.0in}
\end{figure}
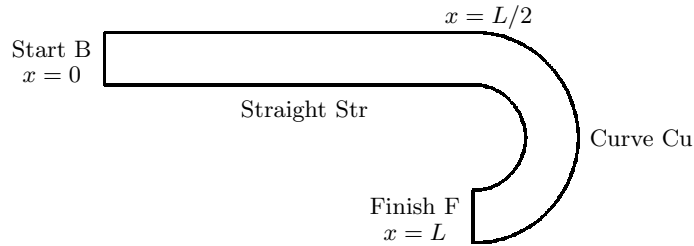

Consider a racing track, pictured in Figure~\ref{chic}, of length $L$ with a car that goes from the start line $B$ to the finish line $F$, the aim is to do this safely and in a short time. We take the transition from straight to curve to be the half way point, $L/2$. The cars have maximum speed 120 $Km/hr$ and cannot go backwards.

%
%
%
%

 As this is a paper more about data rather than control, we choose a very simple control algorithm. We suppose that the speed of the car is given by the setting of a power controller which is marked $[0,120]$, and that there is no need to monitor the speed independently.

\begin{figure}
\begin{center}
\unitlength0.7 mm
\begin{picture}(110,30)(0,22)
\linethickness{0.1mm}
\put(10,30){\line(1,0){100}}
\put(10,25){\makebox(0,0)[cc]{$x=0$}}
\put(42,24.5){\makebox(0,0)[cc]{$\tfrac3{8} L$}}
\put(78,24.5){\makebox(0,0)[cc]{$\tfrac5{8} L$}}
\put(110,25){\makebox(0,0)[cc]{$x=L$}}

\linethickness{0.3mm}
\put(10,30){\line(1,0){32}}
\linethickness{0.3mm}
\multiput(42,30)(0.219,0.12){167}{\line(1,0){0.12}}
\linethickness{0.3mm}
\put(78,50){\line(1,0){32}}
\put(10,30){\makebox(0,0)[cc]{$\bullet$}}
\put(42,30){\makebox(0,0)[cc]{$\bullet$}}
\put(78,30){\makebox(0,0)[cc]{$\bullet$}}
\put(110,30){\makebox(0,0)[cc]{$\bullet$}}

\put(77,42){\makebox(0,0)[cc]{$\phi_{\mathrm{Cu}}(v,x)$}}

\put(0,30){\makebox(0,0)[cc]{$y=0$}}

\put(0,50){\makebox(0,0)[cc]{$y=1$}}


\end{picture}
\caption{The function $\phi_{\mathrm{Cu}}  :  S=[0,120]\times  [0,L]\to [0,1]$.     }
\label{funcphi}
\end{center}
\vspace{-0.25in}
\end{figure}
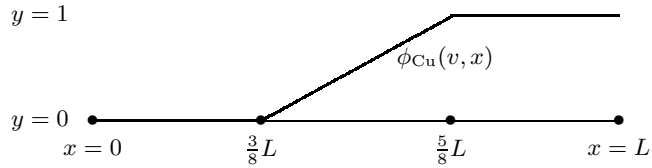

A state space for the whole system is $S=[0,120]\times [0, L]$, which defines velocity and position. We take an open cover of $S$ consisting of open sets representing the straight and curve part of the track:
\[
U_{  \{\mathrm{Str} \}}=[0,120]\times [0,\tfrac34 L)\ ,\
U_{  \{\mathrm{Cu} \}}=[0,120]\times (\tfrac14 L,L]\ ,\
U_{  \{\mathrm{Str},\mathrm{Cu} \}}=[0,120]\times ( \tfrac14 L,\tfrac34 L)\ .
\]
The abstract simplical complex is
$\mathcal{C} = \big\{ \{\mathrm{Str}\}     ,\{\mathrm{Cu}\}   , \{\mathrm{Str}      ,\mathrm{Cu}       \}    \big\}        $ and we specify the partition of unity $\phi :  S=[0,120]\times  [0,L]\to \Delta_\mathcal{C}$ by graphing the
$\phi_{\mathrm{Cu}} $ component (the component taking nonzero values only in $U_{  \{\mathrm{Cu} \}}$, see Definition~\ref{realpu}) in
Figure~\ref{funcphi}. For this choice of function note that $\phi_{\mathrm{Cu}} (v,x)$ is independent of $v$
and only depends on $x\in[0,L]$. Then we have $\phi_{\mathrm{Str}} =1-\phi_{\mathrm{Cu}} $.

\subsection{The data types for the modes}
As the original state space is already given in simple numerical terms, for the data structures we can simply take
the sets themselves, i.e.\ $S_X=U_X$ above.
Recalling \ref{compstr}, in this simple case we can simplify the description of the algebras for the three modes by using a common inherited structure.  We suppose that $\mathcal{A}_X$ contains the datatypes:

\smallskip
\noindent
\begin{tabular}{|c|c|c|c|c|c}\hline  \textbf{Type} &  \textbf{Comment}  &  $X=\{\mathrm{Str}\}$   &  $X=\{\mathrm{Str,Cu}\}$  & $X=\{\mathrm{Cu}\}$ \\ \hline  
$S_{  X } $ &  local state space & $[0,120]\times $     & $[0,120]\times  $    & $[0,120]\times $  \\
& & $ [0,\tfrac34 L)$ & $ (\tfrac14L,\tfrac34 L)$ & $(\tfrac14 L,L]$ \\
 speed${}_X$  &   speed of car in \textit{Km/s} & $[0,120]$   &$[0,120]$   &$[0,120]$     \\
 position${}_X$  &  position of car in $[0,L]$ & $ [0,\tfrac34 L)$ & $ (\tfrac14L,\tfrac34 L)$ & $(\tfrac14 L,L]$  \\
       $\mathcal{C}$          &  $\big\{ \{\mathrm{Cu}\},   \{\mathrm{Str},\mathrm{Cu}\} , \{\mathrm{Str}\} \big\}$  &   &   &   \\
             check       &   \{OK, NotOK\}   & & &    \\
              null          & \{0\}   & & &     \\
               $\Delta_{ \mathcal{C}}$    &  $[0,1] $  &   &  &  \\\hline \end{tabular}
             
              \smallskip
\noindent

%
%
%
              \smallskip
\noindent
In this simple case we avoid special notation for simplicial complexes by identifying $\Delta_{ \mathcal{C}  }$
with the unit interval $[0,1]$, where the vertex $e_{  \mathrm{Str} }$ corresponds to $0\in [0,1]$ and $e_{  \mathrm{Cu} }$ corresponds to $1\in [0,1]$. Thus the position of the system in $\Delta_{ \mathcal{C}  }=[0,1]$ is simply given by the value of $\phi_{\mathrm{Cu}} $ in Figure~\ref{funcphi}.
In addition $\mathcal{A}_X$ contains the following functions, oracles and transfers, given in a general case for mode $X$.
       
       \smallskip
       \noindent      
 \begin{tabular}{|c|c|c|c|}\hline  \textbf{Function} & \textbf{Value}  &  \textbf{Comment} \\\hline  
$\mathrm{updateSpeed}:\mathrm{speed}_X\times S_X\to S_X$ & $\big(s_1,(s_2,p)\big)\mapsto (s_1,p)$    &
update speed    \\
 $\mathrm{updatePosition}:\mathrm{position}_X\times S_X\to S_X$&  $\big(p_1,(s,p_2)\big)\mapsto (s,p_1)$    &  update position   \\
$\mathrm{retrieveSpeed}:   S_X\to \mathrm{speed}_X$  & $(s,p)\mapsto s$    & retrieve speed     \\
$\mathrm{retrievePosition}:  S_X\to \mathrm{position}_X$    & $(s,p)\mapsto p$   &  retrieve position   \\
$\phi_X:S_X \times \mathcal{M} \to [0,1]$ & $(s,\alpha)\mapsto \phi_X(s)_\alpha$  & components of $\phi_X$   \\
 \hline
\textbf{Oracle}    &    &    \\ \hline
    $\mathcal{O}_\mathrm{pos}: \mathrm{null} \to \mathrm{position}_X$    & real world position    & input position    \\
   $\mathcal{O}_\mathrm{power}: \mathrm{speed}_X \to \mathrm{check}$         & OK if successful   & output power (speed)   \\
                & notOK if a problem    &    \\ \hline
                \textbf{Transfer}    &    &    \\ \hline
        $  \mathrm{tran}   : S_X \times \mathcal{C} \to \mathrm{check}$       &   OK if successful  & $  \mathrm{tran}(s,Z)$ transfers     \\
                                   &   NotOK if a problem    &    control to $Z\in \mathcal{C}$ with     \\
                                                                      &       &    initial state $s \in S_X$      \\ \hline \end{tabular}
 
 \medskip The function  $\phi_X:S_X\to \Delta_{ \mathcal{C} } = [0,1]$ is the restriction of $\phi$ to $S_X$.
The oracle $\mathcal{O}_\mathrm{pos}$ is defined by $\mathcal{O}_\mathrm{pos}(0)\in [0,L]$ being the position of the car when the oracle is called (we suppose no delay and no error), and calling the oracle does not affect the \textit{environment} (see the discussion in Example~\ref{posmes}). (We should also have an exception for being out of the specified range, but have chosen not to implement this.)
As the input value to the function is null there is no information transferred to the sensor making the measurement.

The oracle $\mathcal{O}_\mathrm{power}(v)$ is defined by setting the power controller of the car to speed $v\in[0,120]$ (c.f. another actuator or control mechanism in Example~\ref{gearset}). In this case the value returned by the power controller is either OK (the operation has been carried out) or NotOK (a problem occurred).

\subsection{Mode transitions}
Suppose that we are in mode $X\in\mathcal{C}$.
Mode transitions to simpler (subset) modes $Y\subset X$ take priority over moving to more complicated (superset) modes. We shall move to a subset mode on leaving the interior of $\Delta_X$ (see (\ref{inter})) in the simplicial complex.
The only time we expect to make a transition to a subset mode is from $\{  \mathrm{Str} , \mathrm{Cu}    \}$ to $\{\mathrm{Cu}\}$, so this appears only in algorithm $\mathcal{B}_{ \{ \mathrm{Str} , \mathrm{Cu} \} }$ in Section~\ref{algos3}
where we change mode only when $t:=\phi_X(\mathit{state},\mathrm{Cu})$  takes the value 1.

The only time we expect to make a transition to a superset mode is from $\{\mathrm{Str}\}$ to
$\{  \mathrm{Str} , \mathrm{Cu}    \}$. From Section~\ref{funjup} in mode $X=\{\mathrm{Str}\}$ the belief that that mode is doing a good job modelling the system is $\phi_\mathrm{Str}=1-\phi_\mathrm{Cu}$. This has critical values the `comfort level' $0<\kappa_{X} <1$ where we consider moving to a larger (superset) mode
in the subset of $S_X$ and a `panic level' $0<\pi_{X} <\kappa_{X} <1$ where we
consider it urgent to move to a superset mode. To make the transition in plenty of time we set a large value of $\kappa_{   \{\mathrm{Str}\}  } $, say $\kappa_{   \{\mathrm{Str}\}  } =\frac9{10}$. To make sure that we perform the transition before the end of the straight we set
$\pi_{   \{\mathrm{Str}\}  }  =\frac6{10}$. As a result the algorithm $\mathcal{B}_{ \{ \mathrm{Str} \} }$
has the test $\phi_X(\mathit{state},\mathrm{Str})<\frac9{10}$ for mode transfer.

A consequence of the simplicity of this system is that the transition functions themselves are rather boring: all state spaces are just $[0,120]$ cross a suitable subset of $[0,L]$, and all transition functions are inclusion maps on similar sets.\footnote{We could have made things pointlessly more difficult by having distances on the straight and on the curves by measured by different units  (linear v polar; imperial v metric units). Programmers working on real world systems will be familiar with such legacy issues. } We also set all the calculated partition of unity $\phi_X$ to be the abstract partition $\phi$ restricted to $S_X$.
Thus the main interest for the transition functions is checking their domains.
From Definition~\ref{projDom} we have
\begin{align*}
&\mathrm{Dom}\big(   {}_{\{\mathrm{Str}\}}\mathrm{proj}{}_{\{  \mathrm{Str} , \mathrm{Cu}    \}} \big) \supset
\Big\{  \tilde s\in S_{\{  \mathrm{Str} , \mathrm{Cu}    \}}  :  \phi_{   \mathrm{Cu}  }(\tilde s) <
\epsilon_{{\{  \mathrm{Str} , \mathrm{Cu}    \}} \to\{\mathrm{Str}\} }
\Big\} \\ & \quad= [0,120]\times   \Big(\frac14 L, \frac{3 + 2\epsilon_{{\{  \mathrm{Str} , \mathrm{Cu}    \}} \to\{\mathrm{Str}\} }
}{8} L\Big)
\\
&\mathrm{Dom}\big(   {}_{   \{\mathrm{Cu}\}  }\mathrm{proj}{}_{\{  \mathrm{Str} , \mathrm{Cu}    \}} \big) \supset
\Big\{  \tilde s\in S_{\{  \mathrm{Str} , \mathrm{Cu}    \}}  :  \phi_{   \mathrm{Str}  }(\tilde s) <\epsilon_{{\{  \mathrm{Str} , \mathrm{Cu}    \}} \to{   \{\mathrm{Cu}\}  }}\Big\} \\ & \quad= [0,120]\times   \Big(\frac{5 - 2\epsilon_{{\{  \mathrm{Str} , \mathrm{Cu}    \}} \to\{\mathrm{Cu}\} }
}{8} L,\frac34 L\Big)
\end{align*}
From Definition~\ref{incDom} we have
\begin{align*}
& \mathrm{Dom}\big(  {}_{\{  \mathrm{Str} , \mathrm{Cu}    \}}\mathrm{inc}{}_{   \{\mathrm{Str}\}  }    \big) \supset \Big\{  \tilde s\in S_{   \{\mathrm{Str}\}  } : \pi_{   \{\mathrm{Str}\}  }  \le  \phi_{   \{\mathrm{Str}\}  }(\tilde s)    \ \mathrm{and}\
 \phi_{   \{\mathrm{Cu}\}  }  (\tilde s)  >0  \Big\} \\ &\quad
 = [0,120]\times   \Big(\frac38 L, \frac{5-2 \pi_{   \{\mathrm{Str}\}  } }{8}L\Big]
\\
& \mathrm{Dom}\big(  {}_{\{  \mathrm{Str} , \mathrm{Cu}    \}}\mathrm{inc}{}_{   \{\mathrm{Cu}\}  }    \big) \supset \Big\{  \tilde s\in S_{   \{\mathrm{Cu}\}  } : \pi_{   \{\mathrm{Cu}\}  }  \le \phi_{   \{\mathrm{Cu}\}  }(\tilde s)    \ \mathrm{and}\
 \phi_{   \{\mathrm{Str}\}  }  (\tilde s)  >0  \Big\} \\ & \quad
  = [0,120]\times   \Big[ \frac{3+2\pi_{   \{\mathrm{Cu}\}  } }{8}L , \frac 58 L\Big)
\end{align*}

\subsection{The algorithms for the three modes} \label{algos3}
             \noindent
Recall the rough templates proposed in subsection \ref{compstr}.  The algorithm $\mathcal{B}_{ \{ \mathrm{Str} \} }$ is
\\
\\
\noindent\fbox{
\parbox{\textwidth}{%

 {\fontfamily{pcr}\selectfont  \textbf{declaration  }} $\mathit{state}: S_{  \{  \mathrm{Str}  \} }$, $v:\mathrm{speed}_{  \{  \mathrm{Str}  \} }$, $x:\mathrm{position}_{  \{  \mathrm{Str}  \} }$,
$\mathit{checkTransfer,checkSpeed}: \mathrm{check}$

$\mathit{state}:=(0,0)$ \quad \% set presumed beginning state

$x:=\mathcal{O}_\mathrm{pos}(0)$ \quad \% $x$ is now the current position from the environment

$\mathit{state}:=\mathrm{updatePosition}(x,\mathit{state})$  \quad \% update the state with the measured $x$

 $v:=120$

 $\mathit{checkSpeed}:=\mathcal{O}_\mathrm{power}(v)$ \quad \% we choose not to check for exceptions for the power controller
 
 $\mathit{state}:=\mathrm{updateSpeed}(v,\mathit{state})$  \quad \% update the state with our velocity setting (not measured)

 {\fontfamily{pcr}\selectfont  \textbf{loop }}

\quad $x:=\mathcal{O}_\mathrm{pos}(0)$ \quad \% $x$ is now the current position from the environment

\quad     $\mathit{state}:=\mathrm{updatePosition}(x,\mathit{state})$  \quad \% update state with new position

\quad  {\fontfamily{pcr}\selectfont  \textbf{if}} $\phi_X(\mathit{state},\mathrm{Str})<\frac9{10}$
{\fontfamily{pcr}\selectfont  \textbf{then}}
 $ \mathit{checkTransfer}:=\mathrm{tran}(   \mathit{state}, \{  \mathrm{Str},\mathrm{Cu}    \}       )$    
 {\fontfamily{pcr}\selectfont  \textbf{else}} $\mathit{checkTransfer}:=\mathrm{OK}$

%

\quad  {\fontfamily{pcr}\selectfont  \textbf{if}} $\mathit{checkTransfer}=\mathrm{NotOK}$
{\fontfamily{pcr}\selectfont  \textbf{then}}
 $\mathit{checkSpeed}:=\mathcal{O}_\mathrm{power}(0)$ \quad \% if mode transfer fails, stop car

 {\fontfamily{pcr}\selectfont  \textbf{return }}
}
}
\\
\\
\noindent
The algorithm $\mathcal{B}_{ \{ \mathrm{Str},\mathrm{Cu}  \} }$ is
\\
\\
\noindent\fbox{
\parbox{\textwidth}{%

 {\fontfamily{pcr}\selectfont  \textbf{declaration  }}  $\mathit{state}: S_{ \{ \mathrm{Str},\mathrm{Cu}  \} }$,   $v : \mathrm{speed}_{ \{ \mathrm{Str},\mathrm{Cu}  \} }$, $x : \mathrm{position}_{ \{ \mathrm{Str},\mathrm{Cu}  \} }$,
$\mathit{checkTransfer,checkSpeed} :  \mathrm{check}$,  $t :  [0,1]$

 {\fontfamily{pcr}\selectfont  \textbf{input}} $\mathit{state}$  \quad \% the state is transferred from $ \{ \mathrm{Str} \}$.
 
%

 {\fontfamily{pcr}\selectfont  \textbf{loop }}

\quad $x:=\mathcal{O}_\mathrm{pos}(0)$ \quad \% $x$ is now the current position from the environment

\quad     $\mathit{state}:=\mathrm{updatePosition}(x,\mathit{state})$  \quad \% update state with new position

\quad $t:=\phi_X(\mathit{state},\mathrm{Cu})$

\quad $v:=80 \times t+ 120\times(1-t)$

\quad $\mathit{checkSpeed}:=\mathcal{O}_\mathrm{power}(v)$  \quad \% Implement a gradual slow down to 80 $Km/hr$

\quad  {\fontfamily{pcr}\selectfont  \textbf{if}} $t=1$
{\fontfamily{pcr}\selectfont  \textbf{then}}
 $ \mathit{checkTransfer}:=\mathrm{tran}(   \mathit{state}, \{  \mathrm{Cu}    \}       )$    
 {\fontfamily{pcr}\selectfont  \textbf{else}} $\mathit{checkTransfer}:=\mathrm{OK}$

%

\quad  {\fontfamily{pcr}\selectfont  \textbf{if}} $\mathit{checkTransfer}=\mathrm{NotOK}$
{\fontfamily{pcr}\selectfont  \textbf{then}}
 $\mathit{checkSpeed}:=\mathcal{O}_\mathrm{power}(0)$ \quad \% if mode transfer fails, stop car

 {\fontfamily{pcr}\selectfont  \textbf{return }}
}
}
\\
\\
\smallskip\noindent
The algorithm $\mathcal{B}_{ \{ \mathrm{Cu} \} }$ is
\\
\\
\noindent\fbox{
\parbox{\textwidth}{%

 {\fontfamily{pcr}\selectfont  \textbf{declaration  }}  $\mathit{state}: S_{ \{ \mathrm{Cu} \} }$,   $v : \mathrm{speed}_{ \{ \mathrm{Cu} \} }$,
$\mathit{checkSpeed} :  \mathrm{check}$

 {\fontfamily{pcr}\selectfont  \textbf{input}} $\mathit{state}$  \quad \% the state is transferred from $ \{ \mathrm{Str} ,  \mathrm{Cu}  \}$.

 $v:=80$

$\mathit{checkSpeed}:=\mathcal{O}_\mathrm{power}(v)$ \quad \% we choose not to check for exceptions for the power controller

}}

\subsection{A product system and introducing interactions}

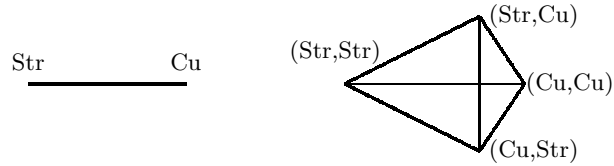
\begin{figure}
\begin{center}
\unitlength 0.6 mm
\begin{picture}(120,35)(0,12)
\linethickness{0.3mm}
\put(5,30){\line(1,0){35}}
\put(5,35){\makebox(0,0)[cc]{Str}}

\put(40,35){\makebox(0,0)[cc]{Cu}}

\linethickness{0.3mm}
\multiput(75,30)(0.24,-0.12){125}{\line(1,0){0.24}}
\linethickness{0.15mm}
\put(75,30){\line(1,0){40}}
\linethickness{0.3mm}
\multiput(105,15)(0.12,0.18){83}{\line(0,1){0.18}}
\linethickness{0.3mm}
\multiput(75,30)(0.24,0.12){125}{\line(1,0){0.24}}
\linethickness{0.3mm}
\multiput(105,45)(0.12,-0.18){83}{\line(0,-1){0.18}}
\linethickness{0.3mm}
\put(105,15){\line(0,1){30}}
\put(73,37){\makebox(0,0)[cc]{(Str,Str)}}

\put(117,45){\makebox(0,0)[cc]{(Str,Cu)}}

\put(117,15){\makebox(0,0)[cc]{(Cu,Str)}}

\put(125,30){\makebox(0,0)[cc]{(Cu,Cu)}}

\end{picture}
\caption{The simplical complex $\mathcal{C}$ for one car (left) and $\mathcal{C}_1\times \mathcal{C}_2$ for two cars (right)}
\label{productcar}
\end{center}
\vspace{-0.0in}
\end{figure}

The simplical complex $\mathcal{C}$ for one car for our track in Figure~\ref{chic} is quite simple, it is the 1-simplex on the left in Figure~\ref{productcar}. To make this more interesting we could consider two cars (car 1 and car 2) on the track.
The reader may remember toy racing tracks where each car had a separate slot, so that the cars could not collide. In this case there is no need for the cars to interact, they can be controlled entirely separately, and then we get the product $\mathcal{C}_1 \times \mathcal{C}_2$ of the simplicial complexes for one car, which is the 3-simplex (solid tetrahedron) on the
 right in Figure~\ref{productcar}. There the vertex $(\mathrm{Str},\mathrm{Cu})$ corresponds to car 1 being in mode $\{\mathrm{Str}\}$ and car 2 being in mode $\{\mathrm{Cu}\}$.
Then $\mathcal{C}_1 \times \mathcal{C}_2$ consists of all subsets of 
$$\big\{  (  \mathrm{S} , \mathrm{S} )  ,(  \mathrm{S} , \mathrm{Cu} )  ,(  \mathrm{Cu} , \mathrm{S} )  ,(  \mathrm{Cu} , \mathrm{Cu} )  
\big\}$$
and the partition of unity $\psi:[0,L]\times [0,L] \to \Delta_{ \mathcal{C}_1 \times \mathcal{C}_2 }$ is given by
\begin{align*}
\psi(x_1,x_2) &= \phi_{\mathrm{Cu}} (x_1)   \,  \phi_{\mathrm{Cu}}  (x_2) \, e_{  (  \mathrm{Cu} , \mathrm{Cu} )     }
+ (1- \phi_{\mathrm{Cu}} (x_1) )  \,  \phi_{\mathrm{Cu}}  (x_2) \, e_{  (  \mathrm{S} , \mathrm{Cu} )     } \\
& \quad + \phi_{\mathrm{Cu}} (x_1)   \, (1- \phi_{\mathrm{Cu}}  (x_2) )\, e_{  (  \mathrm{Cu} , \mathrm{S} )     }
+ (1- \phi_{\mathrm{Cu}} (x_1) )  \, (1- \phi_{\mathrm{Cu}}  (x_2)) \, e_{  (  \mathrm{S} , \mathrm{S} )     }
\end{align*}
For simplicity, here and later, we suppress mention of the velocity in the state space
 and only consider the position coordinates $(x_1,x_2)\in [0,L]\times [0,L]$, where $x_i$ is the position of car $i$.

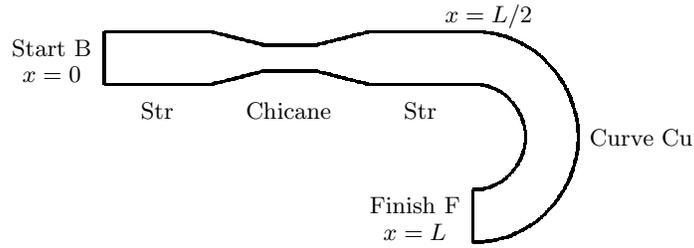
\begin{figure}
\begin{center}
\unitlength 0.7 mm
\begin{picture}(115,40)(10,29)
\linethickness{0.3mm}
\linethickness{0.3mm}
\linethickness{0.3mm}
\linethickness{0.3mm}
\linethickness{0.3mm}
\linethickness{0.3mm}
\linethickness{0.3mm}

\linethickness{0.3mm}
\multiput(90,40)(0.5,0.01){1}{\line(1,0){0.5}}
\multiput(90.5,40.01)(0.5,0.04){1}{\line(1,0){0.5}}
\multiput(91,40.05)(0.49,0.06){1}{\line(1,0){0.49}}
\multiput(91.49,40.11)(0.49,0.09){1}{\line(1,0){0.49}}
\multiput(91.98,40.2)(0.49,0.11){1}{\line(1,0){0.49}}
\multiput(92.47,40.31)(0.48,0.14){1}{\line(1,0){0.48}}
\multiput(92.95,40.44)(0.47,0.16){1}{\line(1,0){0.47}}
\multiput(93.42,40.6)(0.23,0.09){2}{\line(1,0){0.23}}
\multiput(93.88,40.79)(0.23,0.1){2}{\line(1,0){0.23}}
\multiput(94.34,40.99)(0.22,0.11){2}{\line(1,0){0.22}}
\multiput(94.78,41.22)(0.22,0.12){2}{\line(1,0){0.22}}
\multiput(95.21,41.47)(0.21,0.14){2}{\line(1,0){0.21}}
\multiput(95.63,41.74)(0.2,0.15){2}{\line(1,0){0.2}}
\multiput(96.04,42.03)(0.13,0.1){3}{\line(1,0){0.13}}
\multiput(96.43,42.34)(0.12,0.11){3}{\line(1,0){0.12}}
\multiput(96.8,42.67)(0.12,0.12){3}{\line(1,0){0.12}}
\multiput(97.16,43.02)(0.11,0.12){3}{\line(0,1){0.12}}
\multiput(97.5,43.38)(0.11,0.13){3}{\line(0,1){0.13}}
\multiput(97.82,43.77)(0.1,0.13){3}{\line(0,1){0.13}}
\multiput(98.12,44.16)(0.14,0.21){2}{\line(0,1){0.21}}
\multiput(98.4,44.57)(0.13,0.21){2}{\line(0,1){0.21}}
\multiput(98.66,45)(0.12,0.22){2}{\line(0,1){0.22}}
\multiput(98.9,45.44)(0.11,0.22){2}{\line(0,1){0.22}}
\multiput(99.12,45.89)(0.1,0.23){2}{\line(0,1){0.23}}
\multiput(99.31,46.35)(0.17,0.47){1}{\line(0,1){0.47}}
\multiput(99.48,46.82)(0.15,0.48){1}{\line(0,1){0.48}}
\multiput(99.63,47.29)(0.12,0.48){1}{\line(0,1){0.48}}
\multiput(99.75,47.77)(0.1,0.49){1}{\line(0,1){0.49}}
\multiput(99.85,48.26)(0.07,0.49){1}{\line(0,1){0.49}}
\multiput(99.92,48.76)(0.05,0.5){1}{\line(0,1){0.5}}
\multiput(99.97,49.25)(0.02,0.5){1}{\line(0,1){0.5}}
\put(100,49.75){\line(0,1){0.5}}
\multiput(99.97,50.75)(0.02,-0.5){1}{\line(0,-1){0.5}}
\multiput(99.92,51.24)(0.05,-0.5){1}{\line(0,-1){0.5}}
\multiput(99.85,51.74)(0.07,-0.49){1}{\line(0,-1){0.49}}
\multiput(99.75,52.23)(0.1,-0.49){1}{\line(0,-1){0.49}}
\multiput(99.63,52.71)(0.12,-0.48){1}{\line(0,-1){0.48}}
\multiput(99.48,53.18)(0.15,-0.48){1}{\line(0,-1){0.48}}
\multiput(99.31,53.65)(0.17,-0.47){1}{\line(0,-1){0.47}}
\multiput(99.12,54.11)(0.1,-0.23){2}{\line(0,-1){0.23}}
\multiput(98.9,54.56)(0.11,-0.22){2}{\line(0,-1){0.22}}
\multiput(98.66,55)(0.12,-0.22){2}{\line(0,-1){0.22}}
\multiput(98.4,55.43)(0.13,-0.21){2}{\line(0,-1){0.21}}
\multiput(98.12,55.84)(0.14,-0.21){2}{\line(0,-1){0.21}}
\multiput(97.82,56.23)(0.1,-0.13){3}{\line(0,-1){0.13}}
\multiput(97.5,56.62)(0.11,-0.13){3}{\line(0,-1){0.13}}
\multiput(97.16,56.98)(0.11,-0.12){3}{\line(0,-1){0.12}}
\multiput(96.8,57.33)(0.12,-0.12){3}{\line(1,0){0.12}}
\multiput(96.43,57.66)(0.12,-0.11){3}{\line(1,0){0.12}}
\multiput(96.04,57.97)(0.13,-0.1){3}{\line(1,0){0.13}}
\multiput(95.63,58.26)(0.2,-0.15){2}{\line(1,0){0.2}}
\multiput(95.21,58.53)(0.21,-0.14){2}{\line(1,0){0.21}}
\multiput(94.78,58.78)(0.22,-0.12){2}{\line(1,0){0.22}}
\multiput(94.34,59.01)(0.22,-0.11){2}{\line(1,0){0.22}}
\multiput(93.88,59.21)(0.23,-0.1){2}{\line(1,0){0.23}}
\multiput(93.42,59.4)(0.23,-0.09){2}{\line(1,0){0.23}}
\multiput(92.95,59.56)(0.47,-0.16){1}{\line(1,0){0.47}}
\multiput(92.47,59.69)(0.48,-0.14){1}{\line(1,0){0.48}}
\multiput(91.98,59.8)(0.49,-0.11){1}{\line(1,0){0.49}}
\multiput(91.49,59.89)(0.49,-0.09){1}{\line(1,0){0.49}}
\multiput(91,59.95)(0.49,-0.06){1}{\line(1,0){0.49}}
\multiput(90.5,59.99)(0.5,-0.04){1}{\line(1,0){0.5}}
\multiput(90,60)(0.5,-0.01){1}{\line(1,0){0.5}}

\linethickness{0.3mm}
\multiput(90,30)(0.5,0.01){1}{\line(1,0){0.5}}
\multiput(90.5,30.01)(0.5,0.02){1}{\line(1,0){0.5}}
\multiput(91,30.02)(0.5,0.03){1}{\line(1,0){0.5}}
\multiput(91.49,30.06)(0.5,0.04){1}{\line(1,0){0.5}}
\multiput(91.99,30.1)(0.5,0.06){1}{\line(1,0){0.5}}
\multiput(92.49,30.16)(0.49,0.07){1}{\line(1,0){0.49}}
\multiput(92.98,30.22)(0.49,0.08){1}{\line(1,0){0.49}}
\multiput(93.47,30.3)(0.49,0.09){1}{\line(1,0){0.49}}
\multiput(93.96,30.4)(0.49,0.1){1}{\line(1,0){0.49}}
\multiput(94.45,30.5)(0.48,0.12){1}{\line(1,0){0.48}}
\multiput(94.94,30.62)(0.48,0.13){1}{\line(1,0){0.48}}
\multiput(95.42,30.75)(0.48,0.14){1}{\line(1,0){0.48}}
\multiput(95.9,30.89)(0.47,0.15){1}{\line(1,0){0.47}}
\multiput(96.37,31.04)(0.47,0.16){1}{\line(1,0){0.47}}
\multiput(96.84,31.21)(0.47,0.18){1}{\line(1,0){0.47}}
\multiput(97.31,31.38)(0.23,0.09){2}{\line(1,0){0.23}}
\multiput(97.77,31.57)(0.23,0.1){2}{\line(1,0){0.23}}
\multiput(98.23,31.77)(0.23,0.11){2}{\line(1,0){0.23}}
\multiput(98.68,31.98)(0.22,0.11){2}{\line(1,0){0.22}}
\multiput(99.12,32.2)(0.22,0.12){2}{\line(1,0){0.22}}
\multiput(99.57,32.44)(0.22,0.12){2}{\line(1,0){0.22}}
\multiput(100,32.68)(0.21,0.13){2}{\line(1,0){0.21}}
\multiput(100.43,32.93)(0.21,0.13){2}{\line(1,0){0.21}}
\multiput(100.85,33.2)(0.21,0.14){2}{\line(1,0){0.21}}
\multiput(101.27,33.48)(0.2,0.14){2}{\line(1,0){0.2}}
\multiput(101.67,33.76)(0.2,0.15){2}{\line(1,0){0.2}}
\multiput(102.08,34.06)(0.13,0.1){3}{\line(1,0){0.13}}
\multiput(102.47,34.36)(0.13,0.11){3}{\line(1,0){0.13}}
\multiput(102.86,34.68)(0.13,0.11){3}{\line(1,0){0.13}}
\multiput(103.23,35)(0.12,0.11){3}{\line(1,0){0.12}}
\multiput(103.6,35.34)(0.12,0.11){3}{\line(1,0){0.12}}
\multiput(103.96,35.68)(0.12,0.12){3}{\line(0,1){0.12}}
\multiput(104.32,36.04)(0.11,0.12){3}{\line(0,1){0.12}}
\multiput(104.66,36.4)(0.11,0.12){3}{\line(0,1){0.12}}
\multiput(105,36.77)(0.11,0.13){3}{\line(0,1){0.13}}
\multiput(105.32,37.14)(0.11,0.13){3}{\line(0,1){0.13}}
\multiput(105.64,37.53)(0.1,0.13){3}{\line(0,1){0.13}}
\multiput(105.94,37.92)(0.15,0.2){2}{\line(0,1){0.2}}
\multiput(106.24,38.33)(0.14,0.2){2}{\line(0,1){0.2}}
\multiput(106.52,38.73)(0.14,0.21){2}{\line(0,1){0.21}}
\multiput(106.8,39.15)(0.13,0.21){2}{\line(0,1){0.21}}
\multiput(107.07,39.57)(0.13,0.21){2}{\line(0,1){0.21}}
\multiput(107.32,40)(0.12,0.22){2}{\line(0,1){0.22}}
\multiput(107.56,40.43)(0.12,0.22){2}{\line(0,1){0.22}}
\multiput(107.8,40.88)(0.11,0.22){2}{\line(0,1){0.22}}
\multiput(108.02,41.32)(0.11,0.23){2}{\line(0,1){0.23}}
\multiput(108.23,41.77)(0.1,0.23){2}{\line(0,1){0.23}}
\multiput(108.43,42.23)(0.09,0.23){2}{\line(0,1){0.23}}
\multiput(108.62,42.69)(0.18,0.47){1}{\line(0,1){0.47}}
\multiput(108.79,43.16)(0.16,0.47){1}{\line(0,1){0.47}}
\multiput(108.96,43.63)(0.15,0.47){1}{\line(0,1){0.47}}
\multiput(109.11,44.1)(0.14,0.48){1}{\line(0,1){0.48}}
\multiput(109.25,44.58)(0.13,0.48){1}{\line(0,1){0.48}}
\multiput(109.38,45.06)(0.12,0.48){1}{\line(0,1){0.48}}
\multiput(109.5,45.55)(0.1,0.49){1}{\line(0,1){0.49}}
\multiput(109.6,46.04)(0.09,0.49){1}{\line(0,1){0.49}}
\multiput(109.7,46.53)(0.08,0.49){1}{\line(0,1){0.49}}
\multiput(109.78,47.02)(0.07,0.49){1}{\line(0,1){0.49}}
\multiput(109.84,47.51)(0.06,0.5){1}{\line(0,1){0.5}}
\multiput(109.9,48.01)(0.04,0.5){1}{\line(0,1){0.5}}
\multiput(109.94,48.51)(0.03,0.5){1}{\line(0,1){0.5}}
\multiput(109.98,49)(0.02,0.5){1}{\line(0,1){0.5}}
\multiput(109.99,49.5)(0.01,0.5){1}{\line(0,1){0.5}}
\multiput(109.99,50.5)(0.01,-0.5){1}{\line(0,-1){0.5}}
\multiput(109.98,51)(0.02,-0.5){1}{\line(0,-1){0.5}}
\multiput(109.94,51.49)(0.03,-0.5){1}{\line(0,-1){0.5}}
\multiput(109.9,51.99)(0.04,-0.5){1}{\line(0,-1){0.5}}
\multiput(109.84,52.49)(0.06,-0.5){1}{\line(0,-1){0.5}}
\multiput(109.78,52.98)(0.07,-0.49){1}{\line(0,-1){0.49}}
\multiput(109.7,53.47)(0.08,-0.49){1}{\line(0,-1){0.49}}
\multiput(109.6,53.96)(0.09,-0.49){1}{\line(0,-1){0.49}}
\multiput(109.5,54.45)(0.1,-0.49){1}{\line(0,-1){0.49}}
\multiput(109.38,54.94)(0.12,-0.48){1}{\line(0,-1){0.48}}
\multiput(109.25,55.42)(0.13,-0.48){1}{\line(0,-1){0.48}}
\multiput(109.11,55.9)(0.14,-0.48){1}{\line(0,-1){0.48}}
\multiput(108.96,56.37)(0.15,-0.47){1}{\line(0,-1){0.47}}
\multiput(108.79,56.84)(0.16,-0.47){1}{\line(0,-1){0.47}}
\multiput(108.62,57.31)(0.18,-0.47){1}{\line(0,-1){0.47}}
\multiput(108.43,57.77)(0.09,-0.23){2}{\line(0,-1){0.23}}
\multiput(108.23,58.23)(0.1,-0.23){2}{\line(0,-1){0.23}}
\multiput(108.02,58.68)(0.11,-0.23){2}{\line(0,-1){0.23}}
\multiput(107.8,59.12)(0.11,-0.22){2}{\line(0,-1){0.22}}
\multiput(107.56,59.57)(0.12,-0.22){2}{\line(0,-1){0.22}}
\multiput(107.32,60)(0.12,-0.22){2}{\line(0,-1){0.22}}
\multiput(107.07,60.43)(0.13,-0.21){2}{\line(0,-1){0.21}}
\multiput(106.8,60.85)(0.13,-0.21){2}{\line(0,-1){0.21}}
\multiput(106.52,61.27)(0.14,-0.21){2}{\line(0,-1){0.21}}
\multiput(106.24,61.67)(0.14,-0.2){2}{\line(0,-1){0.2}}
\multiput(105.94,62.08)(0.15,-0.2){2}{\line(0,-1){0.2}}
\multiput(105.64,62.47)(0.1,-0.13){3}{\line(0,-1){0.13}}
\multiput(105.32,62.86)(0.11,-0.13){3}{\line(0,-1){0.13}}
\multiput(105,63.23)(0.11,-0.13){3}{\line(0,-1){0.13}}
\multiput(104.66,63.6)(0.11,-0.12){3}{\line(0,-1){0.12}}
\multiput(104.32,63.96)(0.11,-0.12){3}{\line(0,-1){0.12}}
\multiput(103.96,64.32)(0.12,-0.12){3}{\line(0,-1){0.12}}
\multiput(103.6,64.66)(0.12,-0.11){3}{\line(1,0){0.12}}
\multiput(103.23,65)(0.12,-0.11){3}{\line(1,0){0.12}}
\multiput(102.86,65.32)(0.13,-0.11){3}{\line(1,0){0.13}}
\multiput(102.47,65.64)(0.13,-0.11){3}{\line(1,0){0.13}}
\multiput(102.08,65.94)(0.13,-0.1){3}{\line(1,0){0.13}}
\multiput(101.67,66.24)(0.2,-0.15){2}{\line(1,0){0.2}}
\multiput(101.27,66.52)(0.2,-0.14){2}{\line(1,0){0.2}}
\multiput(100.85,66.8)(0.21,-0.14){2}{\line(1,0){0.21}}
\multiput(100.43,67.07)(0.21,-0.13){2}{\line(1,0){0.21}}
\multiput(100,67.32)(0.21,-0.13){2}{\line(1,0){0.21}}
\multiput(99.57,67.56)(0.22,-0.12){2}{\line(1,0){0.22}}
\multiput(99.12,67.8)(0.22,-0.12){2}{\line(1,0){0.22}}
\multiput(98.68,68.02)(0.22,-0.11){2}{\line(1,0){0.22}}
\multiput(98.23,68.23)(0.23,-0.11){2}{\line(1,0){0.23}}
\multiput(97.77,68.43)(0.23,-0.1){2}{\line(1,0){0.23}}
\multiput(97.31,68.62)(0.23,-0.09){2}{\line(1,0){0.23}}
\multiput(96.84,68.79)(0.47,-0.18){1}{\line(1,0){0.47}}
\multiput(96.37,68.96)(0.47,-0.16){1}{\line(1,0){0.47}}
\multiput(95.9,69.11)(0.47,-0.15){1}{\line(1,0){0.47}}
\multiput(95.42,69.25)(0.48,-0.14){1}{\line(1,0){0.48}}
\multiput(94.94,69.38)(0.48,-0.13){1}{\line(1,0){0.48}}
\multiput(94.45,69.5)(0.48,-0.12){1}{\line(1,0){0.48}}
\multiput(93.96,69.6)(0.49,-0.1){1}{\line(1,0){0.49}}
\multiput(93.47,69.7)(0.49,-0.09){1}{\line(1,0){0.49}}
\multiput(92.98,69.78)(0.49,-0.08){1}{\line(1,0){0.49}}
\multiput(92.49,69.84)(0.49,-0.07){1}{\line(1,0){0.49}}
\multiput(91.99,69.9)(0.5,-0.06){1}{\line(1,0){0.5}}
\multiput(91.49,69.94)(0.5,-0.04){1}{\line(1,0){0.5}}
\multiput(91,69.98)(0.5,-0.03){1}{\line(1,0){0.5}}
\multiput(90.5,69.99)(0.5,-0.02){1}{\line(1,0){0.5}}
\multiput(90,70)(0.5,-0.01){1}{\line(1,0){0.5}}

\linethickness{0.3mm}
\put(90,30){\line(0,1){10}}
\put(79,37){\makebox(0,0)[cc]{Finish F}}
\put(79,32){\makebox(0,0)[cc]{$x=L$}}
\put(10,67){\makebox(0,0)[cc]{Start B}}
\put(10,62){\makebox(0,0)[cc]{$x=0$}}

\put(93,73){\makebox(0,0)[cc]{$x=L/2$}}

\put(122,50){\makebox(0,0)[cc]{Curve Cu}}

\put(30,55){\makebox(0,0)[cc]{Str}}
\put(55,55){\makebox(0,0)[cc]{Chicane}}
\put(80,55){\makebox(0,0)[cc]{Str}}



\put(0,30){
\linethickness{0.3mm}
\put(20,30){\line(0,1){10}}
\linethickness{0.3mm}
\put(20,30){\line(1,0){20}}
\linethickness{0.3mm}
\put(20,40){\line(1,0){20}}
\linethickness{0.3mm}
\multiput(40,40)(0.48,-0.12){21}{\line(1,0){0.48}}
\linethickness{0.3mm}
\multiput(40,30)(0.48,0.12){21}{\line(1,0){0.48}}
\linethickness{0.3mm}
\put(50,32.5){\line(1,0){10}}
\linethickness{0.3mm}
\multiput(60,32.5)(0.48,-0.12){21}{\line(1,0){0.48}}
\linethickness{0.3mm}
\put(50,37.5){\line(1,0){10}}
\linethickness{0.3mm}
\multiput(60,37.5)(0.48,0.12){21}{\line(1,0){0.48}}
\linethickness{0.3mm}
\put(70,40){\line(1,0){20}}
\linethickness{0.3mm}
\put(70,30){\line(1,0){20}}
}

\end{picture}

\caption{A track for two racing cars with a chicane}
\label{chic2}
\end{center}
\vspace{-0.0in}
\end{figure}

To introduce interaction between the cars we use a chicane where the track narrows so that the cars can collide, as shown in Figure~\ref{chic2}. Now in the mode
$(\mathrm{Str},\mathrm{Str})$ in the product and only in that mode we have to allow for interaction. In any other position in the 3-simplex we have at least one car either in the curve or in the region where the straight and curve are both in play near position $L/2$, and thus at least one car is beyond the chicane so we have no concerns about a collision.

\begin{figure}
\begin{center}
\unitlength 0.7 mm
\begin{picture}(160,60)(0,14)
\linethickness{0.3mm}
\put(30,50){\line(0,1){10}}
\linethickness{0.3mm}
\put(30,50){\line(1,0){10}}
\linethickness{0.3mm}
\put(40,50){\line(0,1){10}}
\linethickness{0.3mm}
\put(30,60){\line(1,0){10}}
\linethickness{0.3mm}
\put(15,65){\line(1,0){10}}
\linethickness{0.3mm}
\put(25,45){\line(0,1){20}}
\linethickness{0.3mm}
\put(25,45){\line(1,0){20}}
\linethickness{0.3mm}
\put(45,35){\line(0,1){10}}
\linethickness{0.3mm}
\put(15,35){\line(1,0){30}}
\linethickness{0.3mm}
\put(15,35){\line(0,1){30}}
\linethickness{0.1mm}
\multiput(5,25)(0.12,0.12){125}{\line(1,0){0.12}}
\linethickness{0.1mm}
\put(20,40){\line(0,1){27}}
\put(20,67){\vector(0,1){0.12}}
\linethickness{0.1mm}
\multiput(26,25)(0.12,0.15){75}{\line(0,1){0.15}}
\linethickness{0.1mm}
\put(35,36){\line(1,0){13}}
\put(48,36){\vector(1,0){0.12}}
\linethickness{0.1mm}
\multiput(16,25)(0.12,0.13){100}{\line(0,1){0.13}}
\linethickness{0.1mm}
\put(28,38){\line(1,0){20}}
\put(48,38){\vector(1,0){0.12}}
\linethickness{0.1mm}
\multiput(2,43)(0.12,0.12){117}{\line(0,1){0.12}}
\linethickness{0.1mm}
\put(16,57){\line(0,1){10}}
\put(16,67){\vector(0,1){0.12}}
\linethickness{0.1mm}
\multiput(4,35)(0.12,0.12){117}{\line(1,0){0.12}}
\linethickness{0.1mm}
\put(18,49){\line(0,1){18}}
\put(18,67){\vector(0,1){0.12}}
\linethickness{0.1mm}
\multiput(35,59)(0.12,-0.12){33}{\line(1,0){0.12}}
\linethickness{0.1mm}
\multiput(32,57)(0.12,-0.12){42}{\line(1,0){0.12}}
\linethickness{0.1mm}
\multiput(31,53)(0.12,-0.12){17}{\line(1,0){0.12}}
\put(35,64){\makebox(0,0)[cc]{chicane}}

\linethickness{0.3mm}
\put(0,20){\line(1,0){51}}
\put(51,20){\vector(1,0){0.12}}
\linethickness{0.3mm}
\put(0,20){\line(0,1){46}}
\put(0,66){\vector(0,1){0.12}}
\linethickness{0.3mm}
\put(70,20){\line(0,1){40}}
\linethickness{0.3mm}
\put(70,20){\line(1,0){40}}
\linethickness{0.3mm}
\put(110,20){\line(0,1){40}}
\linethickness{0.3mm}
\put(70,60){\line(1,0){40}}
\put(35,15){\makebox(0,0)[cc]{$x_1$}}

\put(-5,60){\makebox(0,0)[cc]{$x_2$}}

\linethickness{0.3mm}
\put(120,60){\line(1,0){40}}
\linethickness{0.3mm}
\put(160,20){\line(0,1){40}}
\linethickness{0.3mm}
\put(120,20){\line(1,0){40}}
\linethickness{0.3mm}
\put(120,20){\line(0,1){40}}
\linethickness{0.3mm}
\put(90,20){\line(0,1){40}}
\linethickness{0.3mm}
\put(120,40){\line(1,0){40}}
\linethickness{0.3mm}
\put(72.5,35){\line(1,0){5}}
\linethickness{0.3mm}
\put(77.5,22.5){\line(0,1){12.5}}
\linethickness{0.3mm}
\put(72.5,22.5){\line(1,0){5}}
\linethickness{0.3mm}
\put(72.5,22.5){\line(0,1){12.5}}
\linethickness{0.3mm}
\put(122.5,27.5){\line(1,0){12.5}}
\linethickness{0.3mm}
\put(135,22.5){\line(0,1){5}}
\linethickness{0.3mm}
\put(122.5,22.5){\line(1,0){12.5}}
\linethickness{0.3mm}
\put(122.5,22.5){\line(0,1){5}}
\put(90,64){\makebox(0,0)[cc]{car 1}}

\put(140,64){\makebox(0,0)[cc]{car 2}}

\put(70,20){\makebox(0,0)[cc]{$\bullet$}}

\put(110,20){\makebox(0,0)[cc]{$\bullet$}}

\put(120,20){\makebox(0,0)[cc]{$\bullet$}}

\put(160,20){\makebox(0,0)[cc]{$\bullet$}}

\put(120,60){\makebox(0,0)[cc]{$\bullet$}}

\put(70,60){\makebox(0,0)[cc]{$\bullet$}}

\put(70,64){\makebox(0,0)[cc]{$(0,L)$}}

\put(120,64){\makebox(0,0)[cc]{$(0,L)$}}

\put(120,15){\makebox(0,0)[cc]{$(0,0)$}}

\put(70,15){\makebox(0,0)[cc]{$(0,0)$}}

\put(110,15){\makebox(0,0)[cc]{$(L,0)$}}

\put(160,15){\makebox(0,0)[cc]{$(L,0)$}}

\put(100,40){\makebox(0,0)[cc]{Cu}}

\put(140,50){\makebox(0,0)[cc]{Cu}}

\put(80,45){\makebox(0,0)[cc]{Str}}

\put(147.5,30){\makebox(0,0)[cc]{Str}}

\put(129,25){\makebox(0,0)[cc]{W}}

\put(75,29){\makebox(0,0)[cc]{W}}

\end{picture}
\caption{avoiding a collision at the chicane}
\label{chioof}
\end{center}
\vspace{-0.25in}
\end{figure}
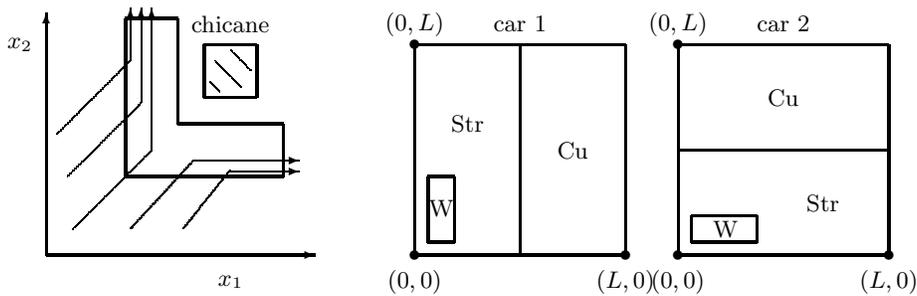

In the $[0,L]\times [0,L]$ picture of the state space the chicane appears as a forbidden square (shaded in the left picture in Figure~\ref{chioof}), where we do not allow both cars to be in the chicane at the same time. As the cars cannot reverse, their paths move up and to the right as time increases. The simplest way to avoid the square is to place a shield downwards and to the left to deflect incoming paths away from the square, and that is what we show in the left picture in Figure~\ref{chioof}, where we have magnified the region around the chicane. This arrowhead shield is made from the union of two overlapping rectangles, which are marked with W in the two rightmost pictures in Figure~\ref{chioof}.

The rightmost pictures in Figure~\ref{chioof} are two covers of the joint state space $ [0,L]\times [0,L]$ (we have suppressed the velocity). The centre picture is the cover for car 1, and the rightmost for car 2. We beg the reader's indulgence for not showing the overlaps between the subsets in the cover, but the pictures were complicated enough already. The dynamics needed to shield the forbidden chicane square are now quite simple. If the path of the joint system in the cover for car $i$ enters the subset W (for Wait), then car $i$ stops, and does not start until the path leaves W. This is illustrated in the outside four paths entering the arrowhead in the leftmost picture. The only complication is what happens when the path enters the intersection of the car 1 and car 2 W subsets, i.e.\ the (W,W) mode of the joint system. Then, according to the previous instruction, both cars will come to a halt. The simplest thing is to have a timer on the cars to restart them, with the timer on car 2 being shorter so that it starts first. This is shown on the middle path entering the arrowhead, where both cars halt for a time when the path changes direction.

\begin{figure}
\begin{center}
\unitlength 0.6 mm
\begin{picture}(95,50)(0,8)
\linethickness{0.3mm}
\multiput(50,30)(0.18,-0.12){167}{\line(1,0){0.18}}
\linethickness{0.15mm}
\put(50,30){\line(1,0){40}}
\linethickness{0.3mm}
\multiput(80,10)(0.12,0.24){83}{\line(0,1){0.24}}
\linethickness{0.3mm}
\multiput(50,30)(0.18,0.12){167}{\line(1,0){0.18}}
\linethickness{0.3mm}
\multiput(80,50)(0.12,-0.24){83}{\line(0,-1){0.24}}
\linethickness{0.3mm}
\put(80,10){\line(0,1){40}}
\linethickness{0.3mm}
\multiput(20,50)(0.18,-0.12){167}{\line(1,0){0.18}}
\linethickness{0.3mm}
\put(20,10){\line(0,1){40}}
\linethickness{0.3mm}
\multiput(20,10)(0.18,0.12){167}{\line(1,0){0.18}}
\linethickness{0.3mm}
\multiput(10,30)(0.12,0.24){83}{\line(0,1){0.24}}
\linethickness{0.3mm}
\multiput(10,30)(0.12,-0.24){83}{\line(0,-1){0.24}}
\linethickness{0.15mm}
\put(10,30){\line(1,0){40}}
\put(50,38){\makebox(0,0)[cc]{(Str,Str)}}

\put(10,50){\makebox(0,0)[cc]{(Str,W)}}

\put(10,10){\makebox(0,0)[cc]{(W,Str)}}

\put(1,30){\makebox(0,0)[cc]{(W,W)}}

\put(91,50){\makebox(0,0)[cc]{(Str,Cu)}}

\put(91,10){\makebox(0,0)[cc]{(Cu,Str)}}

\put(100,30){\makebox(0,0)[cc]{(Cu,Cu)}}

\end{picture}
\caption{The simplicial complex for two cars with a chicane, two solid tetrahedra meeting at a point}
\label{twotet}
\end{center}
\vspace{-0.25in}
\end{figure}
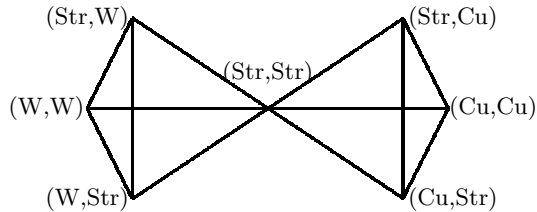

The complete simplicial complex for our two car system is shown in Figure~\ref{twotet}. On the right we have the product simplex from Figure~\ref{productcar}. This has now been altered by changing only the
$(\mathrm{Str},\mathrm{Str})$ mode, and now we join it to the 3-simplex which is the product complex for two copies of the $\{\mathrm{Str},\mathrm{W}\}$ complex. However this second (leftmost) 3-simplex does not simply have the product algorithm (i.e.\ two algorithms for each car running completely independently). Here the algorithm for car $i$ has to track the position of the other car, and we also have the different timing for the car $i$ W algorithms.

If the reader looks looks again at Figure~\ref{chioof} they will see that we have been rather careless, we have no subset to take care of the forbidden square where two cars are simultaneously in the chicane. If that were to happen, the control systems would simply assume that they were both in mode Str, and disaster would likely ensue. However, that cannot happen, as our arrowhead shield means that the cars could never be in that position, is that not so? We note: 
``No plan survives contact with the enemy."\footnote{After Helmuth von Moltke the Elder (1800-1891).}

\section{Concluding remarks}\label{concluding_remarks}

\subsection{Summary}

Our task was to create a mathematical model of a system that operates in a finite number of distinct modes. Our novel idea is to model the algebraic structure of a family of modes by an abstract simplicial complex $\mathcal{C}$. To bring together the modes to create a model of the whole system, we have used the abstract simplicial complex $\mathcal{C}$ to derive appropriate sheaves containing the key components of the modes $X \in \mathcal{C}$, namely: the packages $D_X$ with their state spaces $S_X$, data types $\mathcal{D}_X$ and algorithms $\mathcal{B}_X$
and the maps needed for mode transitions. 

To formally quantify -- and visualise geometrically -- we have used the abstract simplicial complex to make a concrete simplicial complex $\Delta_{\mathcal{C}} \subset \mathbb{R}^n$ containing a simplex $\Delta_X$ for each mode $X \in \mathcal{C}$.  Functions $\phi_X$ that evaluate states, and thresholds $\kappa, \pi$  that trigger transitions, complete the quantificaton. 

To develop the model, we will need a variety of case studies. However, some existing mathematical theories offer ready-made general models to help develop and test a theory. For example, the theory of dynamical systems based upon smooth manifolds is a potential general source of examples  -- see the next subsection \ref{manifold}.

New concepts and properties can also be expected to find a role:  the data types associated with a mode/simplex are computable structures; the interface with a real world environment  can be analysed by the theory of \textit{physical oracles} \cite{Axiomatising};  in dealing with the real world we must allow for issues like precision, errors and delays.

\subsection{Mathematical connections: modes and manifolds}\label{manifold}
In applied mathematics, the state space of many physical systems is often idealised as an $n$-dimensional smooth manifold, which is a topological space which has a family of has local coordinates $(x_1,\dots x_n)\in\R^n$ and allows differentiability (\cite{Arnold1973,Arnold1978}. These local coordinate systems are maps that take a covering of open sets of the space into $\R^n$ and are called charts. In certain situations, each chart can be the basis of a mode in our sense.

Consider a system whose global state space is a manifold $M$. A behaviour of the system can be thought of as a point moving on path in the manifold $M$. A mode transition is the passage of the point from one chart to another.

Let $S_\alpha  \subset \R^n$ to be the set of coordinates for chart $U_\alpha$ and
 $S_\beta  \subset \R^n$ to be the set of coordinates for chart $U_\beta$; let the coordinates be denoted  $x^i$ and  $y^i$ for $1\le i\le n$, respectively. 
 Then we have sets of coordinates  corresponding to the intersection
 $U_\alpha\cap U_\beta$ that are the subsets $I_{\alpha\beta}\subset S_\alpha$ and $I_{\beta\alpha}\subset S_\beta$; and there is a smooth transition function $\tau_{\beta\alpha}:I_{\alpha\beta}\to I_{\beta\alpha}$ with inverse $\tau_{\alpha\beta}: I_{\beta\alpha} \to I_{\alpha\beta}$. 

In terms of our model, now we set 
 $S_{  \{\alpha,\beta \} }=I_{\alpha\beta} \times I_{\beta\alpha} $. 
On moving from $U_\alpha$  into the intersection $U_\alpha\cap U_\beta$ (assuming it is non-empty) we take $(x_1\dots x_n)\in I_{\alpha\beta} \subset S_\alpha$ and copy it into $S_{  \{\alpha,\beta \} }$, and then use the change of coordinate map to calculate the $y^i$s, giving 
$\big((x_1\dots x_n),(y_1\dots y_n)\big)\in S_{  \{\alpha,\beta \} }$.  Note that we only have a partial map from $S_\alpha$ into  $S_{  \{\alpha,\beta \} }$.

\subsection{Widening the scope of the simplicial model of modes}\label{scope}

Another direction for further research is to widen the scope of thinking in terms of modes.   At the heart of the model is a formalisation of decision making using abstract and concrete simplicial complexes to classify and quantify evidence. Decision making permeates computing applications. Many decisions in applications involve weighing up data, arguments, evidence etc.\ and making a judgement. Probability theories have helped analyse decision making since the 18th Century, especially in the birth of actuarial mathematics. More recently, belief theories \cite{TheoryOfEvidence}, many valued logics \cite{MVLog}, neural nets and sundry other machine learning techniques, have provided mathematical theories with which to explore inexact reasoning and making decisions. Guidied by formalisation using an abstract simplicial complex, the scope of the modes could become much wider.
\newline
\newline
\textbf{Human systems:} 
The intuitive ideas about modes, first illustrated and expounded in Section \ref{Decision_making}, can be extended from examples of `hard' physical systems to `soft' human systems. To give a flavour of modes and mode transitions involving people, consider this example:

\begin{exam} \label{man5}
Consider a patient with disease $\alpha$ being monitored in a hospital. We can suppose that $S_{ \{\alpha \} }$ is the information held by the hospital about the patient. 
At some stage the condition of the patient changes to indicate that the patient may be developing disease $\beta $. Monitoring the patient is now in a mode $\{\alpha ,\beta \}$ where tests related to both diseases have to be done, and where complications may arise which would not occur in a patient with a single infection of either disease. The simplest way for the hospital to accommodate the newly required information is simply to make $S_{ \{\alpha ,\beta \} }$ strictly contain $S_{ \{\alpha \} }$. 

Now we take a patient in mode $\{\alpha,\beta\}$ (i.e.,\ presumed to have both conditions $\alpha$ and $\beta$).  To move to mode $\beta$ we require evidence that the patient no longer has condition $\alpha$. In that case we would have a map $S_{\{\alpha,\beta\}} \to S_\beta$ which would retain the medical history, but delete the flags for  current monitoring and medication specific to condition $\alpha$. 
%
%
\end{exam}

\noindent \textbf{Typologies for modes:} We can use a simplicial map from the simplicial complex $\mathcal{C}$ of modes to another simplicial complex $\mathcal{E}$ to classify certain aspects of the system. One example is for $\mathcal{E}$ to represent security clearance, and the simplicial map then maps access modes to security clearances. Recall that partially ordered sets have appeared in models of security and information flow \cite{dennInfo}. Different simplicial maps could be used to classify other behaviours of the system or restrict access to oracles, for example in a social system distinguishing between managerial hierarchies and the authority to rewrite regulations. In this manner modes could inherit behaviour from multiple classifying maps in a transparent fashion specified in the system design stage.

\bibliographystyle{compj}

\end{document}